\documentclass[iop]{emulateapj}
\usepackage{graphicx}
\usepackage{aas_macros}
\usepackage {amsmath}
\usepackage{amssymb}
\usepackage{graphics}
\usepackage{wrapfig}
\usepackage{soul}
\usepackage{natbib}
\usepackage{array}
\usepackage[export]{adjustbox}
\usepackage{xcolor}

\newcolumntype{P}[1]{>{\centering\arraybackslash}p{#1}}
\newcolumntype{M}[1]{>{\centering\arraybackslash}m{#1}}

\newcommand{\um}{\ensuremath{{\mu}m}}

\newcommand{\ox}{[{\ion{O}{3}}]\;}
\newcommand{\bluox}{[\ion{O}{3}]~$\lambda$4960\;}
\newcommand{\redox}{[\ion{O}{3}]~$\lambda$5008\;}
\newcommand{\nii}{[{\ion{N}{2}}]\;}
\newcommand{\blun}{[\ion{N}{2}]~$\lambda$6550\;}
\newcommand{\redn}{[\ion{N}{2}]~$\lambda$6585\;}
\newcommand{\hb}{{H$\beta$}\;}
\newcommand{\ha}{{H$\alpha$}\;}
\newcommand{\ybpt}{[{\ion{O}{3}}]/{{H$\beta$}\;}}
\newcommand{\xbpt}{[{\ion{N}{2}}]/{{H$\alpha$}\;}}
\newcommand{\lxbpt}{[{\ion{N}{2}}]/{{H$\alpha$}}}
\newcommand{\lybpt}{[{\ion{O}{3}}]/{{H$\beta$}}}
\newcommand{\lo}{$L_\mathrm{[OIII]}$\;}
\newcommand{\lx}{$L_\mathrm{X}$\;}
\newcommand{\lamo}{$\lambda_\mathrm{[OIII]}$\;}

\newcommand{\lamxx}{$\lambda_\mathrm{X}$\;}
\newcommand{\lhx}{$L_\mathrm{X(2-10\ keV)}$\;}

\newcommand{\lratio}{{$\log(L_\mathrm{X}$}/{$L_\mathrm{[OIII]})$\;}}
\newcommand{\sfratio}{SFR/SFR${_\mathrm{MS}}$\;} 

\shorttitle{ AGN MULTI-WAVELENGTH IDENTIFICATION}
\shortauthors{Azadi et al.}
 \slugcomment{Accepted to ApJ}

\begin{document}

\title{THE MOSDEF SURVEY: AGN MULTI-WAVELENGTH IDENTIFICATION, SELECTION BIASES AND HOST GALAXY PROPERTIES}

\author{
Mojegan Azadi \altaffilmark{1}, 
Alison L. Coil \altaffilmark{1},
James Aird \altaffilmark{2},
Naveen Reddy \altaffilmark{3},
Alice Shapley \altaffilmark{4},
William R. Freeman \altaffilmark{3},
Mariska Kriek \altaffilmark{5},
Gene C. K. Leung \altaffilmark{1}, 
Bahram Mobasher \altaffilmark{3},
Sedona H. Price \altaffilmark{5},
Ryan L. Sanders \altaffilmark{4},
Irene Shivaei \altaffilmark{3},
Brian Siana  \altaffilmark{3}
}
\altaffiltext{1}{Center for Astrophysics and Space Sciences, Department of Physics, University of California, 9500 Gilman Dr., La Jolla, San Diego, CA 92093, USA}
\altaffiltext{2}{Institute of Astronomy, University of Cambridge, Madingley Road, Cambridge CB3 0HA, UK}
\altaffiltext{3}{Department of Physics and Astronomy, University of California, Riverside, 900 University Avenue, Riverside, CA, 92521, USA}
\altaffiltext{4}{Department of Physics and Astronomy, University of California, Los Angeles, 430 Portola Plaza, Los Angeles, CA 90095, USA}
\altaffiltext{5}{Astronomy Department, University of California at Berkeley, Berkeley, CA 94720, USA}

\begin{abstract} 
We present results from the MOSFIRE Deep Evolution Field (MOSDEF) survey on the identification, selection biases, and host galaxy properties of 55 X-ray, IR and optically-selected active galactic nuclei (AGN) at $1.4 < z < 3.8$. We obtain rest-frame optical spectra of galaxies and AGN and use the BPT diagram to identify optical AGN. We examine the uniqueness and overlap of the AGN identified at different wavelengths. There is a strong bias against identifying AGN at any wavelength in low mass galaxies, and an additional bias against identifying IR AGN in the most massive galaxies. AGN hosts span a wide range of star formation rate (SFR), similar to inactive galaxies once stellar mass selection effects are accounted for. However, we find (at $\sim 2-3\sigma$ significance) that IR AGN are in less dusty galaxies with relatively higher SFR and optical AGN in dusty galaxies with relatively lower SFR. X-ray AGN selection does not display a bias with host galaxy SFR. These results are consistent with those from larger studies at lower redshifts. Within star-forming galaxies, once selection biases are accounted for, we find AGN in galaxies with similar physical properties as inactive galaxies, with no evidence for AGN activity in particular types of galaxies. This is consistent with AGN being fueled stochastically in any star-forming host galaxy. We do not detect a significant correlation between SFR and AGN luminosity for individual AGN hosts, which may indicate the timescale difference between the growth of galaxies and their supermassive black holes.

\end{abstract}


\section{Introduction}
\label{sec:intro}

Active galactic nuclei (AGN) are the result of accretion of gas and dust onto the supermassive black holes (SMBHs) at the centers of galaxies and have been investigated in numerous studies during the past two decades \citep[for a recent review of black hole growth see][]{alexander2012drives}. Various observational results have shown evidence for a global connection between the growth of SMBHs and the galaxies in which they live. For example, the relatively tight correlation between the SMBH mass and the bulge stellar mass \citep[e.g.][]{magorrian1998demography} or bulge velocity dispersion \citep[e.g.][]{ferrarese2000fundamental,gebhardt2000relationship,kormendy2011supermassive} supports the idea of a close connection between the growth of SMBHs and their host galaxies. In addition, the similar evolution of the SMBH accretion rate density and star formation rate (SFR) density with redshift indicates a global connection between AGN activity and the formation of stars in galaxies \citep[e.g.][]{boyle1998cosmological,silverman2008luminosity,aird2010evolution,assef2011mid}. However, the details of the coeval growth of galaxies and their SMBHs are not well understood.  

Accretion onto SMBHs releases a tremendous amount of energy, and thus AGN produce significant radiation at X-ray, ultraviolet (UV), optical, infrared (IR), and radio wavelengths. Different studies have used emission at one or more of these wavelengths to identify AGN and subsequently investigate the nature of their host galaxies \citep[e.g.][]{kauffmann2003,Goulding2009,kauffmann2009,aird2012primus,mendez2013primus,Azadi2015,Cowley2016,Harrison2016kmos}.

One of the most reliable methods of identifying AGN is X-ray imaging from deep surveys carried out with the \textit{XMM-Newton}, \textit{Chandra} and, more recently, \textit{NuSTAR} telescopes \citep[for a recent review see][]{brandt2015}. The X-ray emission from AGN is strong enough to outshine the X-ray light associated with intense star formation activity and penetrate regions with high hydrogen column density (up to $N_{H}\approx 10^{23-24}$ cm$^{-2}$). Thus, hard-band (2--10 keV) X-ray selection is sensitive to both unabsorbed and moderately absorbed AGN, and is relatively unaffected by host galaxy dilution. 

However, X-ray emission is strongly absorbed by Compton thick regions with hydrogen column density of $N_{H}> 1.5 \times 10^{24}$ cm$^{-2}$ \citep[e.g.][]{DellaCeca2008,comastri2011,brightman2014}. Studies of local and non-local AGN samples estimate that 10--50\% of the entire AGN population are Compton thick \citep[e.g.][]{Akylas2009,Vignali2010,Alexander2011,Lanzuisi2015,Buchner2015}, demonstrating that such heavily obscured sources represent a sizable fraction of the full AGN population. Also X-ray identification is not successful in identifying less powerful AGN that are accreting at very low rates \citep[e.g.][]{Gilli2007,aird2012primus} or AGN residing in less massive galaxies \citep [e.g.][]{mendez2013primus}.

For heavily obscured AGN that cannot be recovered by X-ray imaging, identification at other wavelengths may be used. The obscuring dust absorbs the UV and optical radiation from the central engine and re-emits thermal radiation at mid-IR (MIR) wavelengths \citep [e.g][]{Neugebauer1979,Rieke1981}. AGN can thus show a red power-law at MIR wavelengths in their spectral energy distributions (SED) \citep [e.g][]{alonso2006,Donley2007,donley2012,Mateos2012}, which can be identified with imaging from the Infrared Array Camera \citep[IRAC;][]{fazio2004} on \textit{Spitzer} or with the \textit{Wide-field Infrared Survey Explorer} \citep[WISE;][]{Wright2010}. Dust heated by AGN is warmer than dust heated by star formation \citep [e.g.][]{donley2012}, which allows AGN to be distinguished from normal star-forming galaxies at these wavelengths. 

Different selection techniques have been proposed to separate AGN from the galaxy population in MIR color-color space \citep[e.g][]{lacy2004,Stern2005,Assef2010,Messias2012,donley2012,Mateos2012}. 
These methods can identify heavily obscured X-ray and optical AGN \citep[e.g.][]{daddi2007,Fiore2008} as well as luminous AGN, regardless of the obscuration and viewing angle \citep[e.g.][]{hao2011}.
However, studies show that samples of AGN based on any of these IR selection techniques suffer from selection biases such that they mainly identify luminous AGN \citep[e.g][]{mendez2013primus}.

AGN also produce significant radiation at optical wavelengths. Broad optical emission lines (FWHM $>$ 2000 km s$^{-1}$) in unobscured AGN and narrow optical emission lines in obscured AGN, which arise from gas located several hundred parsec away from the SMBH (therefore suffering only from moderate obscuration \citep[e.g.][]{keel1994,kauffmann2003}), can provide detailed information about the central SMBH. At low redshifts, optical diagnostics such as the ``BPT diagram" \citep[e.g.][]{baldwin1981}, which shows the optical emission line ratios of \xbpt versus \ybpt, have been widely used to identify AGN. Various studies based on Sloan Digital Sky Survey (SDSS) data indicate that AGN and star-forming galaxies form distinct sequences on the BPT diagram with some overlap \citep[e.g.][for a recent review see \citealt{Heckman2014}]{kauffmann2003,Kewley2006,kauffmann2009}. One of the greatest advantages of this technique is that it can identify less powerful AGN with low accretion rates that might be obscured at other wavelengths. 

Despite its advantages, there are various issues with the BPT diagnostics. The narrow optical emission lines can suffer from significant extinction due to the dust in the galaxy. Also at higher redshifts it is more difficult to detect the required lines for the BPT diagram at high signal-to-noise ratio (as the optical emission lines shift to the near-IR wavelengths where the terrestrial background is higher). With AGN and star formation both being sources of optical emission lines, disentangling the contributions from each of these phenomena can be  another challenge \citep[e.g.][]{kauffmann2009,wild2010,tanaka2011}.
Furthermore, using the BPT diagram at higher redshifts may require re-calibration of the lines separating AGN from the star-forming sequence \citep[e.g.][]{coil2015,shapley2015}.

The selection biases in each identification method indicate that using a single waveband cannot recover the full population of AGN \citep[e.g][]{hickox2009host,Juneau2011,mendez2013primus,Trump2013,Goulding2014}. Many studies have investigated properties of AGN host galaxies using multi-wavelength data at low and moderate redshifts \citep[e.g][]{klesman2012,mendez2013primus,Goulding2014,Sartori2015}. To obtain a better understanding of the properties of AGN host galaxies requires detailed information at $z \sim 1-3$, the epoch of peak of AGN activity; however, current samples at these redshifts are relatively small.

Studying the host galaxies of AGN---revealing the types of galaxies that tend to host AGN---can provide important insights into the physical mechanisms that trigger AGN activity. Furthermore, we can assess whether AGN appear to have an impact on the galaxies that they live in, altering their properties compared to the overall galaxy population. 
Numerous studies have investigated the position of AGN in the color-magnitude diagram, as well as star formation activity, stellar mass, stellar age or colors of AGN host galaxies at different redshifts to investigate the impact of AGN activity on their host galaxies  \citep[e.g.][]{kauffmann2009,schawinski2011hst,aird2012primus,rosario2013nuclear,gerorgakakis2014,hernan2014,rosario2015}. It is known that galaxies in the optical color-magnitude diagram show a strong bimodal behavior and can generally be divided into two populations: the blue cloud of star-forming galaxies and the red sequence with mainly passive, quiescent galaxies \citep[e.g.][]{blanton2003galaxy,bell2003optical,baldry2004quantifying}. Early studies found that the majority of AGN lie on the red sequence \citep[e.g][]{nandra2007aegis,hickox2009host} and concluded that AGN feedback may be shutting down star formation in their host galaxies.
More recent studies, however, highlight the importance of stellar-mass dependent selection effects \citep[e.g][]{silverman2009environments,xue2010color,cardamone2010dust, aird2012primus,Hainline2012}. In fact, most of these studies find that in a sample with matched stellar mass host galaxies, AGN are equally likely to be found in any host population.

Using far-IR (FIR) and sub-mm measurements from \textit{Herschel} and \textit{ALMA}, many recent studies find that AGN predominantly live in star-forming galaxies \citep[e.g.][]{mullaney2012goods,santini2012enhanced,harrison2012no,rosario2013nuclear,Mullaney2015}. 
However, \textit{Herschel} observations are biased towards FIR bright galaxies.
\cite{aird2012primus} and \cite{Azadi2015} use samples of moderate luminosity AGN and stellar mass complete galaxies  from the PRIMUS redshift survey to show that AGN reside in both the quiescent and star-forming galaxy populations, although galaxies that are forming stars are 2-3 times more likely to host an AGN. While these studies find evidence of enhanced star formation activity in AGN hosts (compared to the inactive galaxies with a similar mass distribution), uncertainties in estimates of SFR (e.g. due to the depth of the observations, the effects of dust reddening, or corrections for AGN contamination), as well as the various selection biases inherent to AGN samples, limit our understanding of the connection between black hole growth and the growth of galaxies, especially at high redshifts.

A number of studies have investigated whether there is a correlation between the SFR and AGN activity in individual galaxies \citep[see, e.g.,][and references therein]{Azadi2015}. Tracking star formation activity only in circumnuclear regions ($r <1$kpc), \cite{diamond2012relationship} find evidence of a correlation between the luminosity of nearby Seyferts and their nuclear star formation and conclude that these process are related in the very central regions of galaxies \citep[see also ][]{Lamassa2013}. Tracking galaxy-wide star formation, studies of moderate luminosity AGN typically find no significant correlation between SFR and AGN activity \citep[e.g.][]{rosario2012mean,mullaney2012goods,Stanley2015,Azadi2015}, while studies of the most luminous AGN find a positive trend that could be driven by major mergers \citep[e.g.][]{rosario2012mean,dai2015,Bernhard2016}. However, rapid variability in the AGN luminosity can result in scatter, consequently washing out the intrinsic correlation between AGN luminosity and SFR. Therefore considering the average AGN luminosity for samples of galaxies of a fixed SFR, instead of the luminosities of individual AGN, may be more appropriate for exploring the relationships between AGN activity and star formation \citep[e.g.][]{hickox2014black}. In fact, studies investigating  average AGN luminosity in bins of SFR find a positive correlation between AGN luminosity and SFR \citep[e.g.][]{chen2013correlation,Azadi2015,dai2015}.

Due to the reliability of  X-ray AGN identification, the majority of the studies discussed above use only X-ray imaging to identify AGN and subsequently assess the properties of their host galaxies using multi-wavelength data. 
As noted above, MIR imaging and optical rest-frame spectroscopy can also identify AGN that may be obscured at other wavelengths. 
Optical spectra in particular can also provide detailed information about the gas, dust and stellar populations of the AGN host galaxies. Until recently such detailed information has been only available at low redshifts, but with the advent of multi-object near-infrared (NIR) spectrographs such as KMOS \citep{Sharples2013} and MOSFIRE \citep{mclean2012}, this information can be obtained at high redshift as well \citep[e.g.][]{Trump2013,Genzel2014,coil2015,Harrison2016kmos}.

In this paper we use rest-frame optical spectra from the MOSFIRE Deep Evolution Field (MOSDEF) survey \citep{kriek2015} taken with the MOSFIRE multi-object NIR spectrograph on the Keck I telescope to investigate AGN identification at multiple wavelengths and their host galaxy properties at $z\sim $1.37 -- 3.80. Considering data from the first season of the MOSDEF survey, \cite{coil2015} found that while the BPT diagram works well for identifying optical AGN at $z\sim2$, it cannot provide a complete sample of AGN, as it suffers from biases against low mass and/or high specific star formation rate (sSFR) host galaxies. In this paper, with a larger dataset from the first two years of the MOSDEF survey, we identify optical AGN using the BPT diagram and use additional AGN samples selected \emph{a priori} based on X-ray and IR imaging data to investigate the selection biases from each identification method. We explore the host galaxy properties of these AGN and investigate the relation between star formation and AGN luminosity in our sample at the 
epoch of the peak of both AGN and galaxy growth.

The outline of the paper is as follows: in Section \ref{sec:data} we describe the X-ray and IR data used to identify AGN in MOSDEF, along with the method used for measuring the optical emission line ratios that is used for identifying optical AGN. In this section we also provide information about stellar mass and SFR estimates in our sample. In Section \ref{sec:results} we present our results on the AGN host galaxy properties and the relation between AGN activity and SFR at $z\sim2$. In Section \ref{sec:discussion} we discuss our results, and we conclude in Section \ref{sec:summary}. Throughout the paper we adopt a flat cosmology with $\Omega_{\Lambda}$ =0.7 and $H_0$=72 km s$^{-1}$Mpc$^{-1}$.

\section{Sample and Measurements}
\label{sec:data}

In this study, we use multi-wavelength data from the MOSDEF survey to investigate AGN host galaxy properties at $z \sim 2$. We use X-ray imaging data from \textit{Chandra}, IR imaging data from \textit{Spitzer-IRAC}, and rest-frame optical spectra obtained with the MOSFIRE spectrograph at Keck Observatory to identify AGN. We describe these datasets below, as well as our methods for fitting the optical emission lines in our spectra and for estimating stellar masses and SFRs of AGN host galaxies by fitting their SEDs.

%%%%%%%%%%%%%%%%%%%%%%%%%%%%%%%%%%%%%%%%%%%%%%%%%%%%%%%%%%%%%%

\subsection{The MOSDEF Survey}
\label{sec:mosdef}

In this study, we use spectroscopic data from the MOSDEF survey \citep{kriek2015}. This survey uses the MOSFIRE spectrograph \citep{mclean2012} on the 10~m Keck I telescope. MOSFIRE provides wavelength coverage from 0.97 to 2.40 $\mu$m with a spectral resolution of R = 3400, 3300, 3650, and 3600 respectively in the Y, J, H, and K bands. MOSDEF observations cover a total area of 500 arcmin${^2}$ in three extragalactic fields:  COSMOS, GOODS-N, and EGS from the CANDELS survey \citep{grogin2011,koekemoer2011} in areas with 3D-\textit{Hubble Space Telescope} (\textit{HST}) grism survey \citep{brammer2012} coverage. Along with \textit{HST}  imaging, there is extensive multi-wavelength ancillary data from other telescopes including \textit{Chandra, Spitzer} and \textit{Herschel} for MOSDEF targets. In this paper we use data from the first two years of the survey.  During this time observations were also taken in the GOODS-S and UDS fields, in addition to the main survey fields above, and these data are included here.

MOSDEF targets span a wide range of redshift from $1.37 < z < 3.80$ and when completed, the survey will include $\sim$1500 galaxies and AGN. The targets are chosen from three distinct redshift intervals ($1.37<z< 1.70$, $2.09 < z<2.61$ and $2.95 < z <3.80$) to ensure that the rest-frame optical emission lines fall within windows of atmospheric transmission. Sources in MOSDEF are targeted down to limiting \textit{HST} /WFC3 F160W magnitudes of 24.0, 24.5, and 25.0, respectively, at $z\sim$ 1.5, 2.3, and 3.4 using the 3D-\textit{HST} photometric catalogs \citep{skelton2014}. Target priorities in MOSDEF are determined by their brightness and redshift information, with brighter sources and those with more secure prior redshift determinations given higher weights. AGN identified in advance via X-ray or IR imaging are also given higher targeting weights. In data from the first two years of the MOSDEF survey, which we use here, we identify 482 galaxies and 55 AGN. Detailed information about the MOSDEF AGN sample is provided below. The full details of the survey, data reduction and analysis are presented in \cite{kriek2015}. 

%%%%%%%%%%%%%%%%%%%%%%%%%%%%%%%%%%%%%%%%%%%%%%%%%%%%%%%%%%%%%%

\subsection{X-ray AGN Selection}
\label{sec:xray}

The X-ray AGN in our sample were identified prior to MOSDEF targeting using the \textit{Chandra} imaging in our fields, which had a depth of 4 Ms in GOODS-S, 2 Ms in GOODS-N, 800 ks in EGS and 160 ks in COSMOS (at the time of MOSDEF target selection). We use catalogs generated in a consistent manner as described by \cite{laird2009} and \cite{nandra2015} (see also \citealt{georgakakis2014,Aird2015}). 
Our adopted Poisson false probability detection threshold ($<4\times10^{-6}$) 
corresponds to reaching hard band flux ($f_\mathrm{2-10\ keV}$) limits (over $>$10\% of the area) of $1.6 \times 10^{-16}$, $2.8 \times 10^{-16}$, $5.0 \times 10^{-16}$ and $1.8 \times 10^{-15}$ in the GOODS-S, GOODS-N, EGS and COSMOS fields, respectively (although we note that the depths of the \textit{Chandra} imaging varies substantially within a field).

 We use the likelihood ratio technique to identify reliable optical, NIR and IRAC counterparts to the X-ray sources (as described in \citealt{Aird2015}; see also \citealt{sutherland1992,Ciliegi2002,Brusa2007,laird2009}). For sources with multiple counterparts, we choose the match with the highest likelihood ratio. Finally, we match our catalog of X-ray counterparts to the 3D-\textit{HST} catalogs used for MOSDEF targeting and find the closest 3D-\textit{HST} match within 1$^{\prime\prime}$.

For our X-ray AGN, we estimate the rest-frame 2--10 keV X-ray luminosities based on the hard (2--7 keV) observed flux or, if the source is not detected in the hard band, the soft (0.5-2 keV) observed flux. We  assume a simple power-law spectrum including only Galactic absorption with a photon index of $\Gamma$ = 1.9. 
We do not correct our X-ray luminosities for intrinsic absorption effects (local to the AGN). 
At $ z\sim 2.3$, only about $10\%$ of the observed X-ray flux is suppressed at a column density of $10^{23}$ cm$^{-2}$. Therefore, our estimates of the X-ray luminosity are accurate and larger absorption column densities ($N_{H}> 10^{23}$ cm$^{-2}$) are required to significantly suppress the observed flux at $z\sim2$.

In total, there are 28 X-ray AGN in the current MOSDEF sample, 22 of which are detected in the hard band with X-ray luminosities ($L_\mathrm{X(2-10\ keV)}$) within the range of 10$^{43}$ to 10$^{45}$ erg s$^{-1}$. Given the relatively high X-ray luminosities of these sources, we do not impose any luminosity cut on the X-ray AGN sample.

\begin{figure}[ht!]
\includegraphics[width=0.485\textwidth]{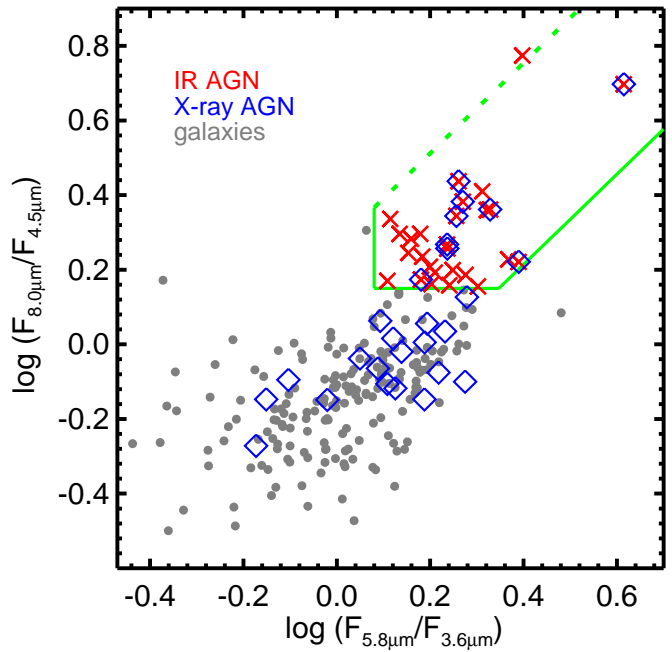}
\caption{IRAC color-color space for MOSDEF sources, where criteria from \citet{donley2012} defined in equations \ref{eq2} to \ref{eq4} in the text are shown in green. The gray points show MOSDEF galaxies, and the blue diamonds are X-ray AGN identified in section \ref{sec:xray}. The red points show the sources that fall inside the Donley-defined wedge and therefore are selected as IR AGN.} 

\label{fig:iragn}
\end{figure}

\subsection{IR AGN Selection}
\label{sec:IR}

Using hard X-ray (2--10 keV) detections ensures that our sample is not strongly biased against moderately obscured ($N_{H}\sim10^{22-24}$ cm$^{-2}$) AGN, however, hard X-ray radiation cannot penetrate Compton-thick regions with heavy obscuration ($N_{H}> 10^{24}$ cm$^{-2}$). In heavily obscured AGN, the high energy nuclear emission is absorbed and reprocessed by dust near the AGN and re-radiated at MIR wavelengths. This re-radiated emission from dust results in MIR imaging also being useful in identifying AGN.

Several MIR AGN selection techniques have been proposed using data from the IRAC \citep{fazio2004} on the \textit{Spitzer} space telescope \citep[e.g.][]{lacy2004,Stern2005,alonso2006,donley2008,donley2012}. Depending on the depth of the data and the redshift of interest, some of these criteria can suffer from contamination from star-forming galaxies mis-identified as AGN. In starburst galaxies or galaxies with older stellar populations and/or higher dust extinction the stellar bump can mimic the power-law emission from AGN.

The \cite{Stern2005} criteria, which were empirically derived from shallow, wide-area \textit{Spitzer} data, have been shown to be unreliable at various redshifts ($z\sim 1$ and $z\gtrsim 2.5$) \citep[e.g.][]{Georgantopoulos2008,donley2012,mendez2013primus}.
Here, using IRAC fluxes from v4.1 of the 3D-\textit{HST} catalogs \citep{skelton2014}, after removing X-ray AGN, we find that $\sim$10\% of the full MOSDEF sample falls within the \cite{Stern2005} selection region (only $\sim$38\%  of these sources are IR AGN using the \citealt{donley2012} criteria), indicating that the \cite{Stern2005} selection may be contaminated by star-forming galaxies at the redshift and depth of our survey.

\cite{donley2012} provide more reliable IR AGN identification criteria, which are designed to limit the contamination from star-forming galaxies, especially at high redshift. The robustness of this selection technique in identifying IR AGN from star-forming galaxies is confirmed by \cite{mendez2013primus}, using a large faint galaxy sample at intermediate redshifts. In the \cite{donley2012} selection criteria, objects are required to be detected in all four IRAC bands and satisfy the following criteria in IRAC color-color space:

\begin{eqnarray}
  x={\rm log_{10}}\left( \frac{f_{\rm 5.8 \um}}{f_{\rm 3.6 \um}} \right), \quad
  y={\rm log_{10}}\left( \frac{f_{\rm 8.0 \um}}{f_{\rm 4.5 \um}} \right)
\end{eqnarray}

\begin{align}
  &x\ge 0.08 \textrm{~ and ~} y \ge 0.15 \textrm{~ and ~} \label{eq2}  \\ 
  &y\ge (1.21\times{x})-0.27 \textrm{~ and ~} \label{eq3}  \\
  &y\le (1.21\times{x})+0.27 \textrm{~ and ~} \label{eq4} \\
  &f_{\rm 4.5 \um} > f_{\rm 3.6 \um} \textrm{~ and ~}  \label{eq5} \\
  &f_{\rm 5.8 \um} >  f_{\rm 4.5 \um} \textrm{~ and ~} \label{eq6}  \\
  &f_{\rm 8.0 \um} > f_{\rm 5.8 \um} \label{eq7} 
\end{align}

In MOSDEF we identify IR AGN using these criteria, with some slight modifications. In addition to the detection in all four IRAC bands, we set a flux limit in each band that corresponds to a S/N respectively in channels 1 to 4 of 3, 3, 2.4, and 2.1 (see \citealt{mendez2013primus} for an explanation of how these limits are derived).

Figure \ref{fig:iragn} shows the relevant IRAC color-color space of ${\log(f_{8.0\um}/f_{4.5\um})}$ versus ${\log(f_{5.8\um}/f_{3.6\um})}$. The green solid and dashed lines indicate the criteria defined in equations \ref{eq2} to \ref{eq4}, and enclose an area referred to as the ``Donley wedge.'' The gray points are galaxies in the MOSDEF sample and the blue diamonds are X-ray AGN identified in section \ref{sec:xray}. The red points show the sources that fall inside the Donley wedge and therefore are selected as IR AGN. There is one red source above the dashed line that satisfies all the above criteria except for equation \ref{eq4}. The IRAC photometry for this source indicates a sharp increase in the flux from channel 1 to 4, which is common only in IR AGN. Since none of the galaxies in MOSDEF lie close to the dashed line, we relax the upper bound on Donley wedge, showing it with a dashed line in Figure \ref{fig:iragn} and including this red source in our IR AGN sample.
There are also three sources inside the Donley wedge that satisfy all the above criteria except for one of the equations \ref{eq5},  \ref{eq6} or \ref{eq7}, which require a strictly increasing, red power-law SED in the IRAC bands. We relax these three criteria such that any source consistent with an increasing red power-law {\it within the 1$\sigma$ errors on the IRAC photometry} is classified as an IR AGN. We note that 32\% of the X-ray AGN lie inside the Donley wedge, which indicates that they are identified as AGN based on both the X-ray and IRAC imaging data. 
In total we identify 27 IR AGN in the current MOSDEF sample, 9 of which are also detected in X-rays.

%%%%%%%%%%%%%%%%%%%%%%%%%%%%%%%%%%%%%%%%%%%%%%%%%%%%%%%%%%%%%%

\begin{figure}[ht!]
\includegraphics[width=0.46\textwidth]{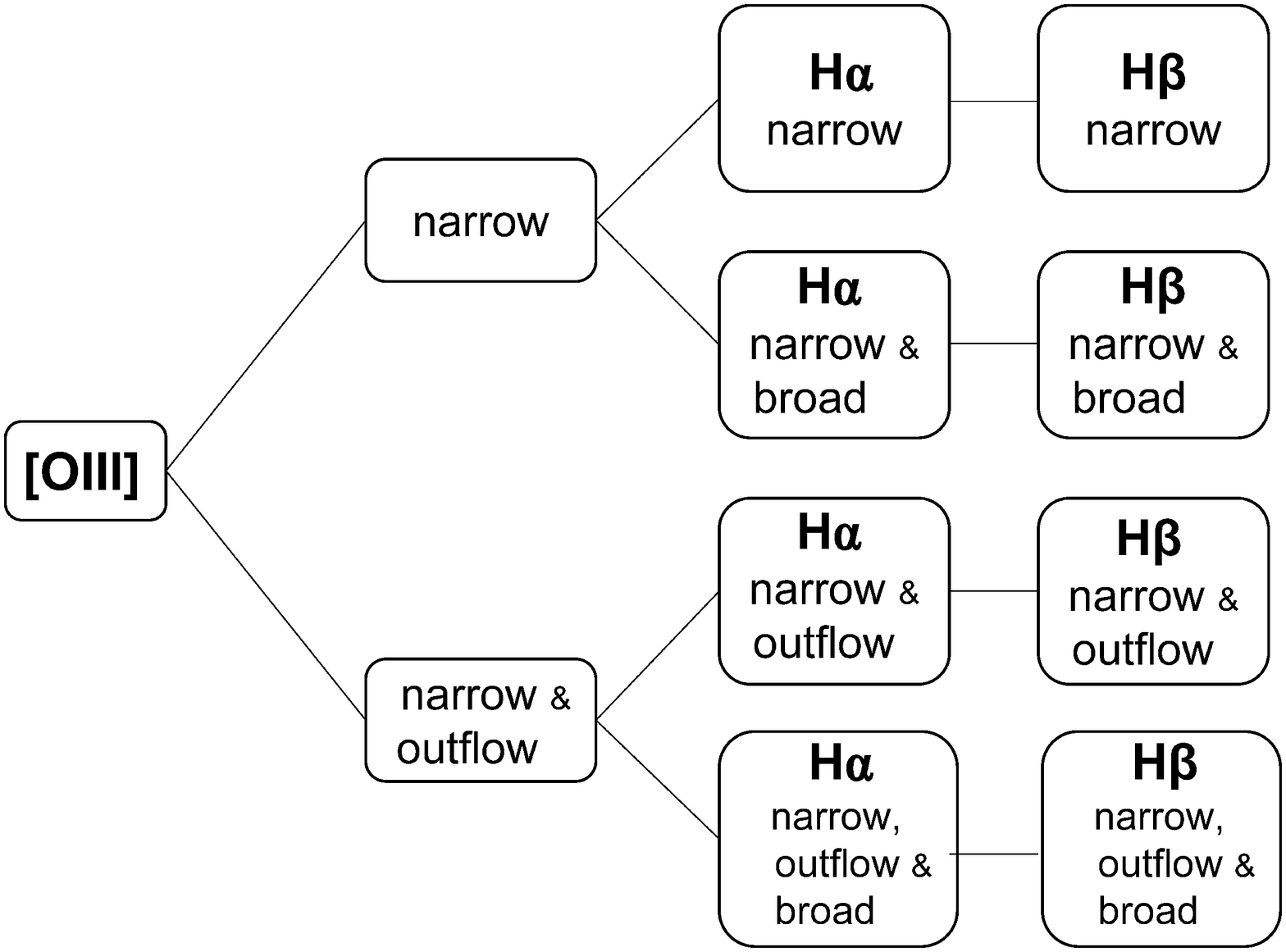}
\caption{A flowchart representing the logic of our Gaussian emission line fitting procedure: In each step we evaluate any change in $\chi^{2}$ to establish whether the additional components results in an improved fit (at the 99\% confidence level).
We first evaluate the fit to \ox to determine whether an additional outflow component is required. We then evaluate whether the \ha fit requires a broad component (also including an outflow component if required by \ox). Lastly, we fit \hb with the same components as \ha.
}
\label{fig:flowchart}
\end{figure}

\subsection{Spectroscopic Data and Optical AGN Selection}
\label{sec:optagn}

The MOSDEF survey has obtained spectroscopic data using the MOSFIRE spectrograph, which enables us to detect emission lines at rest wavelengths of 3700--7000\AA \; for our sources. These data enable us to use optical diagnostics such as the BPT diagram \citep{baldwin1981,veilleux1987} to identify AGN that may not be detectable with X-ray or IR imaging data. 
In this section we describe the emission line fitting procedure we used for measuring the flux of each line, and the AGN that are identified with optical diagnostics.

%%%%%%%%%%%%%%%%%%%%%%%%%%%%%%%%%%%%%%%%%%%%%%%%%%%%%%%%%%%%%%

\subsubsection{Optical Emission Line Flux Measurements}
\label{sec:fit}

In order to use optical diagnostics to identify AGN in our sample, we need to measure the \hb, \ox, \ha, and \nii emission lines. We fit Gaussian functions to the observed lines using the \texttt{MPFIT} nonlinear least-square fitting function in IDL \citep{markwardt2009non}, using the error spectra to determine the errors on the fit. 
We simultaneously fit \redox with \bluox, and \ha with \blun and \redn. We fix the continuum to be flat, with no slope, and allow up to 0.15\% freedom to the centroid of the expected narrow and broad lines wavelength. We require the same physical component (i.e., narrow line, broad line) to have the same FWHM and velocity offset in each line, as determined from the line with the highest S/N. We fix the spacing between the \bluox and \redox forbidden lines and fix their relative normalizations to a ratio of 1:3 (as well as for the \blun and \redn lines). For each source we consider four different models for the emission line profiles, with various components, as described below.

\begin{figure*}[ht!]
\includegraphics[width=0.95\textwidth]{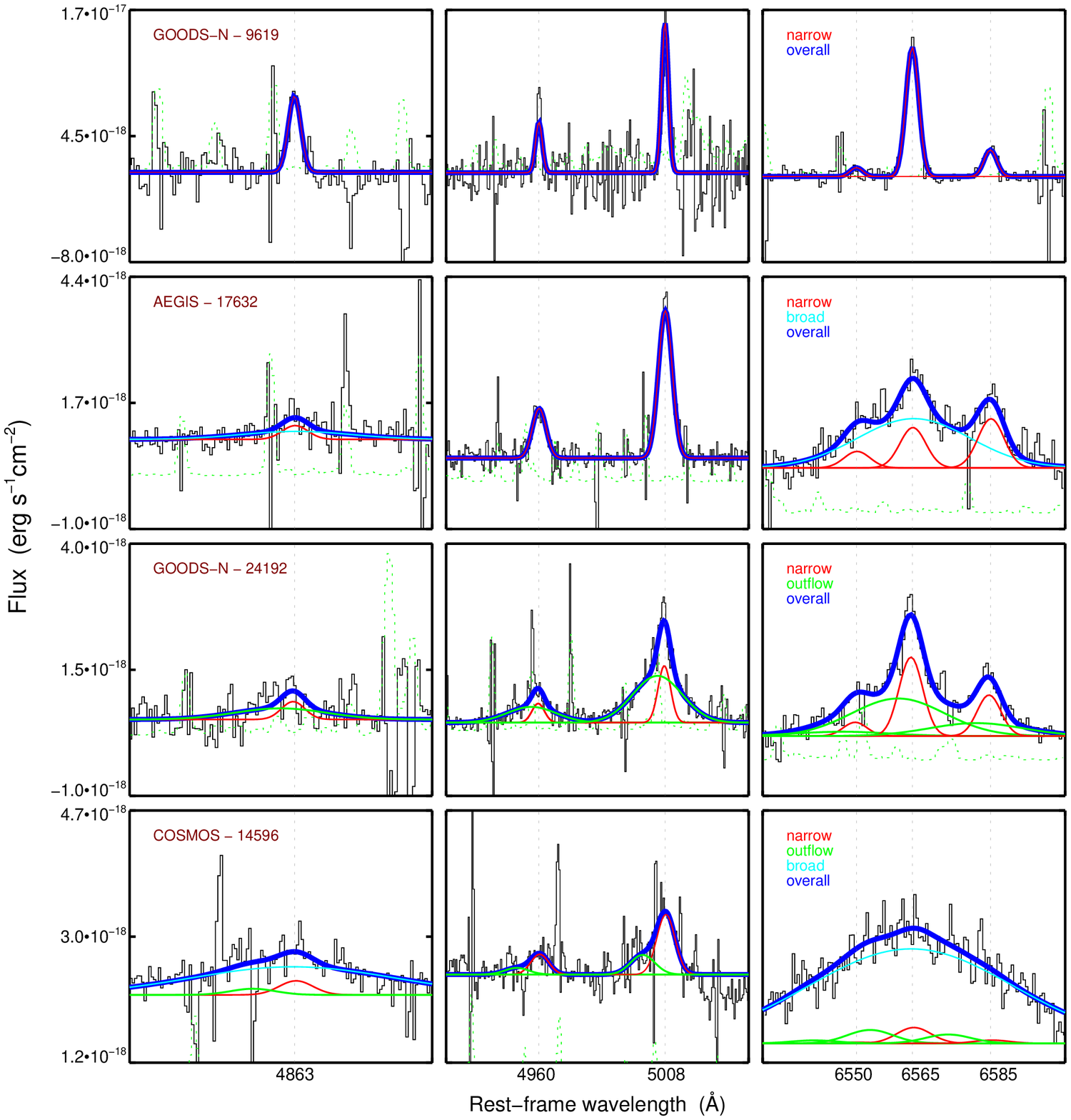}
\caption{Examples of MOSDEF spectra and the multiple-component fitting procedure for four AGN in our sample (the IDs are from v4.1 of the 3D-\textit{HST} catalogs). The overall fit is shown in blue, while the individual Gaussian components are shown in different colors: red for the narrow component, green for an outflow, and cyan for a broad component.  The error spectra are shown with dotted green lines, while vertical dotted gray lines show the rest frame wavelength of the emission lines we fit. The top row shows an AGN where only a narrow Gaussian component is required to all of the lines. The second row shows an AGN that requires an additional broad component to \hb and \ha. The third row shows an AGN with a clear outflow component, detected in the \ox and \ha lines. The fourth row shows an AGN that requires both an outflow component (seen in \redox) and a broad component. Sources that have a broad \ha and/or \hb component with the S/N $>3$
are excluded from our analysis in Section \ref{sec:results}, as we can not reliably probe their host galaxy properties.}
\label{fig:fits}
\end{figure*}

In model 1, we fit each line with a single Gaussian function with the same centroid velocity and FWHM which is required to be $<$2000 km s$^{-1}$, representing the narrow emission line component. In model 2, we fit each line with a narrow Gaussian component and additionally for \hb and \ha include a broad Gaussian component (FWHM $>$2000 km $s^{-1}$) representing emission from the AGN broad-line region. In model 3, we fit each line with two Gaussian components, one narrow line component as above and a second component with a FWHM $<$2000 km s$^{-1}$ and a 
negative velocity offset ($-500 < v < 0$ km s$^{-1}$) relative to the narrow line component, representing an outflow that could be driven by an AGN (Leung et al. 2016 in preparation). The FWHM and centroid velocity of the outflow component is fixed to be identical for all of the emission lines. In model 4, we fit each line with narrow and outflow components, and additionally allow for a broad component with an identical FWHM ($>$2000 km s$^{-1}$) to the \hb and \ha lines.

In each model we evaluate any change in  reduced $\chi^{2}$ to establish whether the additional components result in an improved fit at confidence level of 99\%. Figure \ref{fig:flowchart} is a flowchart that shows the order of our fitting procedure. We first evaluate the reduced $\chi^{2}$  from model 1 to model 3 for the \ox line to see if the additional outflow component has improved the fit. If so, we then evaluate the \ha fit from model 3 to 4 to see if an additional broad component improves the fit. If the \ox line is well fit with only a narrow component, we then compare the \ha fit in model 1 with model 2 to see whether the additional broad component improves the fit.

\begin{figure*}[t]
\centering
\includegraphics[width=0.8\textwidth]{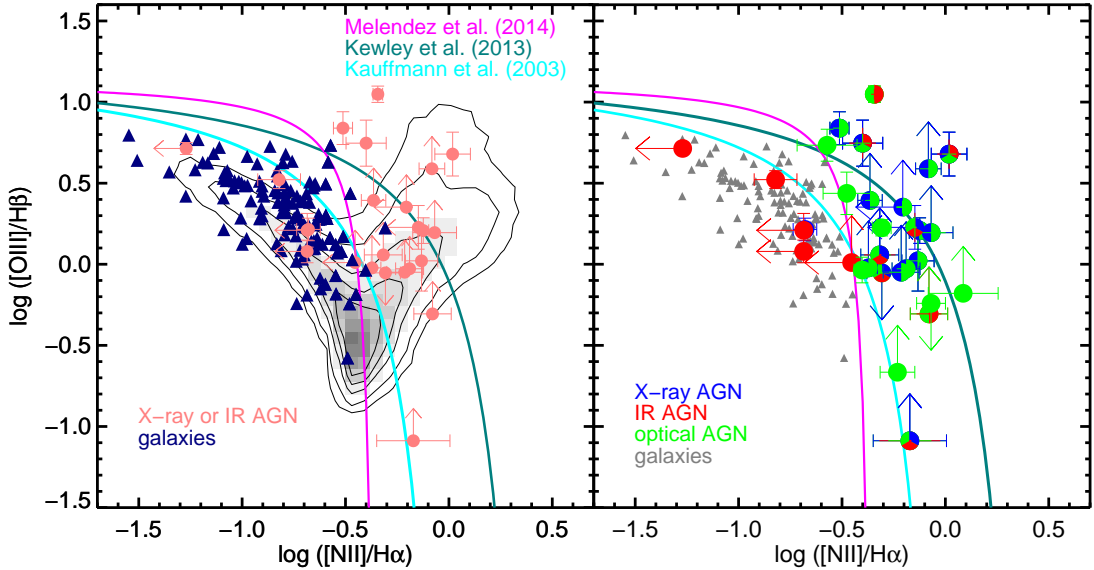} 
\caption{The BPT diagram for MOSDEF AGN and galaxies. \textit{Left:} Pink circles show AGN identified with X-ray and/or IR imaging, with arrows showing $3\sigma$ limits for AGN for which all four relevant optical emission lines are not detected in MOSDEF. Dark blue triangles show MOSDEF galaxies (with $>3\sigma$ detections of all four relevant optical emission lines). The contours and grayscale show the location of SDSS sources in this plane. The cyan line is the empirical criteria from \cite{kauffmann2003} that indicates the division between star-forming galaxies and AGN in SDSS. The dark green line is the theoretical prediction from \cite{kewley2013} of the maximum line ratios expected for starburst galaxies, in the absence of an AGN. The magenta line is the theoretical prediction for identifying AGN by \cite{melendez2014}.
Here we use this line to identify optical AGN in MOSDEF, as this line most cleanly separates known X-ray and/or IR AGN from galaxies in our sample.
\textit{Right:} Similar to the left panel but here MOSDEF AGN 
are shown using different colors that indicate the wavelength used to identify
them as AGN. X-ray AGN are shown in blue, IR AGN are shown in red, and
optical AGN (identified as lying above the \cite{melendez2014} line) 
are shown in green. The AGN that are identified using more than one wavelength 
method are shown with multiple colors. The MOSDEF galaxies with $>3\sigma$ detections  of all four of the necessary emission lines are shown in gray.}
\label{fig:bpt}
\end{figure*}

To find the minimum width of the Gaussian components, we identify sky lines in each wavelength filter for each source, fit them with a single Gaussian component, and use their average width as the minimum width of the narrow line fits. This sets the minimum width for the velocity dispersion of 2.5 \AA \; in the observed H band and 3.5 \AA \;  in the observed K band, which at $z\sim 2.3$ corresponds to $\sim$ 45 km s$^{-1}$  and $\sim$ 48 km s$^{-1}$ respectively in H and K bands.
The minimum width for the broad Gaussian component is set by the upper limit on the FWHM ($\mathrm<$ 2000 km s$^{-1}$) of the narrow lines.

Figure \ref{fig:fits} illustrates examples of MOSDEF spectra for four different AGN in our sample, with the best fit result of our multiple-component fitting procedure. The overall fit is shown in blue, and the individual components are shown in different colors. The top panel shows an AGN where a narrow Gaussian component is the best fit to all the lines. The second panel shows an AGN with a narrow component for all lines and an additional broad component to \hb and \ha. The third panel shows an AGN with a significant outflow component for both the \ox and \ha lines, in addition to the narrow component. In the last panel in addition to the outflow component (seen in the \redox line), the \ha fit improves with an underlying broad component.

In our spectral fits, we find that four AGN (identified at both X-ray and IR wavelengths) require a broad \ha and/or \hb component with the S/N $>3$. Since we can not reliably probe the host galaxy properties of these AGN, we exclude them for the analysis in  Section \ref{sec:results}. We keep these sources in our sample for the purpose of identifying optical AGN below in Section \ref{sec:optagn}, but we do not consider the contributions from the broad optical emission lines.

We note that for all sources where the best fit included an outflow component, we visually inspected the \textit{HST}  imaging to determine whether they might be merger candidates; in these cases the ``outflow'' component might not be from an outflow, but from a merger event.  We identified six sources (ID = 28202, 26028, 22299, 16339, 9183, 3146 in v4.1 of 3D-\textit{HST}) with two potential nuclei that indicate that the host galaxies of these AGN may be undergoing merger events. 
Since it is not clear which of the two components has the detected AGN, and since the photometry of the two components will often be blended for these sources, 
 we exclude them from the analysis of AGN spectral or host galaxy properties in this paper. Of the 28 X-ray AGN in our sample, four are merger candidates, while from the 27 IR AGN, two are merger candidates.

%%%%%%%%%%%%%%%%%%%%%%%%%%%%%%%%%%%%%%%%%%%%%%%%%%%%%%%%%%%%%%

\subsubsection{Optical AGN Sample}
\label{sec:optagn}

We use the fluxes from our line fitting procedure to measure the \xbpt and \ybpt line ratios, required to place sources on the BPT diagram. 
We do not include broad components in these fluxes.
We correct the flux of the narrow components of the \hb and \ha lines for underlying stellar absorption, using the best-fit SED models to the multi-wavelength photometry \citep[for more details see][]{reddy2015}. This correction results in an average change of $\sim$ 0.01 and $\sim$ 0.06 dex respectively in log(\lxbpt) and log(\lybpt) line ratios on the BPT diagram and thus has a small effect on our sample.

Figure \ref{fig:bpt} shows the BPT diagram for MOSDEF AGN and galaxies. In the left panel, AGN identified at either X-ray or IR wavelengths are shown with pink circles; the remaining MOSDEF sources (with $>3\sigma$ detections of all four of the necessary emission lines) are shown with blue triangles. We note that there are galaxies with $<3\sigma$ detections that are not shown in this panel. For the X-ray and IR selected AGN, since they are already identified as AGN with reliable methods we require a significant ($>3\sigma$)
detection of at least one of the two lines required for each ratio.  If the other line is not significantly detected, we use the $3\sigma$ limit on the flux to determine the limit on the line ratio (indicated by the arrows in Figure \ref{fig:bpt}). For the 40 X-ray and/or IR detected AGN in our sample, we are able to place 24 AGN on the BPT diagram (including those with limits).

For comparison, we show the location of SDSS sources with contours and grayscale in this panel. The cyan line is the empirical demarcation from \cite{kauffmann2003} that shows the division between star-forming galaxies and AGN in SDSS. The dark cyan line is the theoretical prediction from \cite{kewley2013} of the maximum line ratios expected in starburst galaxies, in the absence of an AGN. We also show in magenta the theoretical prediction for the lowest line ratios allowed by AGN of \cite{melendez2014}.

\begin{figure}[]
\centering
\vspace{0.8cm}
\includegraphics[width=0.35 \textwidth]{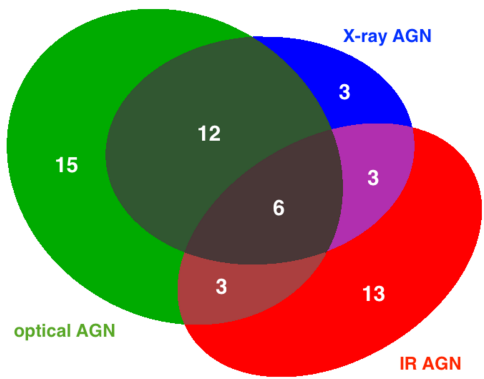}
\vspace{0.55cm}
\caption{A Venn diagram showing the number of AGN in our sample identified at different wavelengths; the full sample contains 55 AGN (and 482 galaxies). The overlapping regions show the number of AGN selected at multiple wavelengths. This diagram shows our detected AGN sample and is not corrected for observational biases such as the depth of the data at each wavelength.}
\label{fig:venn}
\end{figure}

Sources that lie between the Kauffmann et al. and Kewley et al. lines are often referred as ``composite" sources, in that their line ratios can have contributions from both star formation and AGN activity. For sources to the right of the \cite{kewley2013} line, such high line ratios can only be due to AGN activity \citep[e.g.][]{shapley2015}. We note that the galaxies in our sample lie above the main locus of star-forming galaxies in SDSS  in log(\lybpt) and/or log(\lxbpt) (by a median offset of  $\sim$ 0.1 dex, see  \citealt{coil2015,shapley2015}). This offset has been seen in other studies of high redshift galaxies \citep[e.g.][]{yabe2012,masters2014,newman2014,Steidel2014}. Applying a luminosity limit to the SDSS sample that is comparable to the limit for high redshift galaxies reduces this offset somewhat \citep{juneau2014} but as shown in \cite{coil2015} the offset does not completely disappear.

Figure \ref{fig:bpt} shows that the \cite{kauffmann2003} line may not be as reliable as a means of separating galaxy and AGN populations at $z\sim2$ as at $z\sim0$. On the other hand, using the \cite{kewley2013} line, which demarcates a pure sample of AGN, results in a highly incomplete and restricted AGN sample. In fact, the majority of the X-ray or IR detected AGN in the MOSDEF sample lie below the \cite{kewley2013} line. As discussed in \cite{coil2015}, the \cite{melendez2014} line can be used to more reliably separate galaxies and AGN in MOSDEF. The left panel of Figure \ref{fig:bpt} shows that the vast majority of the X-ray and IR detected AGN in our sample lie above this line, while the majority of the other MOSDEF sources lie below this line. Therefore in this study we identify all sources with line ratios that are above this line as optical AGN.

\begin{figure}[t]
\includegraphics[width=0.41\textwidth]{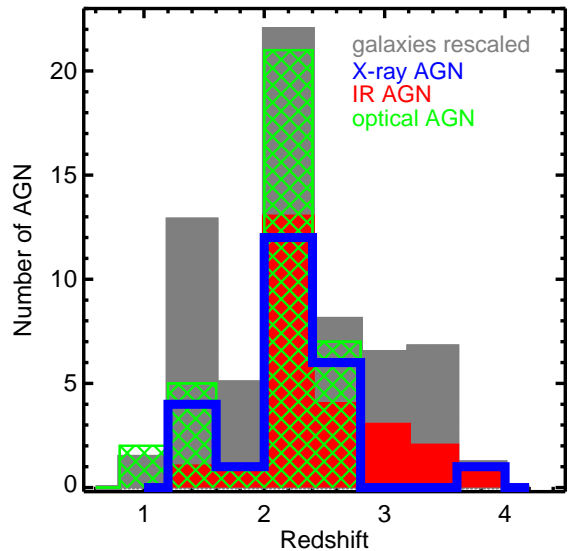}
\caption{The redshift distribution for the 55 AGN and 482 galaxies in the MOSDEF sample. The distribution for each identification technique is shown with a different color.  AGN identified at multiple wavelengths are counted in each distribution and can therefore be represented multiple times in this figure.}
\label{fig:histz}
\end{figure}

The right panel of Figure \ref{fig:bpt} shows the BPT diagram for MOSDEF sources, using the above classification of AGN at different wavelengths. For AGN that are identified using multiple wavelengths, we use multiple colors based on their identification methods. MOSDEF galaxies with at least 3$\sigma$ detections in all four lines are shown with gray triangles. We emphasize that in addition to the four sources above the \cite{melendez2014} line in the left panel of Figure \ref{fig:bpt} which we identify as optical-only AGN, there are 11 sources with 3$\sigma$ detections in both \ha and \redn without reliable \hb and \redox detections. We also classify these sources as optical AGN due to their high log(\lxbpt), which is greater than -0.3.

Using a UV-selected galaxy sample, \cite{Steidel2014} find a larger offset for star-forming galaxies in the BPT diagram, compared with local samples. Using the \cite{melendez2014} line to identify optical AGN in their sample would lead to a larger AGN fraction than in the MOSDEF sample and may lead to more contamination by star-forming galaxies in their sample. Instead, a more strict criterion of log(\lxbpt) $> -0.3$ could be used. In the MOSDEF sample, however, using this cut results in only four sources being removed from the optical AGN sample and therefore does not change any of our conclusions. Since the \cite{melendez2014} line provides the cleanest separation between known AGN and galaxies in the MOSDEF sample, we use it here. We also note that it is unlikely that the integrated light of many of our sources to be dominated by shocks, which could potentially also move sources above the \cite{melendez2014} line.

For the sample used here, there are 40 X-ray and/or IR AGN,
 of which 21 are also identified as optical AGN. 
In addition, there are 15 sources classified as optical AGN that are only identified as AGN through optical diagnostics. In total, there are 55 AGN in the MOSDEF sample. Figure \ref{fig:venn} is a Venn diagram illustrating the number of AGN identified at different wavelengths in our sample. We use ellipses instead of circles in the Venn diagram, so that the areas are proportional to the number of AGN identified with each method.

Figure \ref{fig:histz} illustrates the redshift distribution for AGN and galaxies in our sample. Each color represents the redshift distribution for a different identification technique, and as a single AGN can be detected at multiple wavelengths it can contribute to more than one histogram in this figure. While our AGN span a wide range of redshifts, they strongly peak at $z\sim2$, due to the MOSDEF target selection. We note that we cannot identify optical AGN at $z>2.6$, as at these redshifts the \ha and \redn lines fall beyond the observed 
wavelength coverage of our MOSFIRE spectra.

%%%%%%%%%%%%%%%%%%%%%%%%%%%%%%%%%%%%%%%%%%%%%%%%%%%%%%%%%%%%%%

\subsection{Stellar Mass and SFR Estimates}
\label{sec:sed}

To measure the SFR and stellar mass of the AGN host galaxies, we use SED fitting, which is a widely adopted method for estimating physical properties of galaxies. We use the FAST stellar population fitting code \citep{kriek2009} with the multi-wavelength photometry from 3D-\textit{HST} \citep{skelton2014} and the MOSDEF spectroscopic redshifts. We adopt the \cite{conroy2009} Flexible Stellar Population Synthesis (FSPS) models, along with a \cite{chabrier2003} stellar initial mass function (IMF), the \cite{calzetti2000} dust reddening curve and use a fixed solar metallicity. We consider delayed exponentially declining star formation histories, $\Psi \propto t$ exp $(\frac{-t}{\tau})$, where $t$ is the time since the onset of star formation and $\tau$ is the characteristic star formation timescale and is within the range of $ 0.1 < \tau < 10 $ Gyr.  FAST searches over a grid of models and uses $\chi^{2}$ fitting to determine the best fit solution \citep{kriek2009}. 

There is one AGN (identified at both X-ray and optical wavelengths) in our sample that is not fully spatially covered by the CANDELS imaging in the COSMOS field, such that the photometry is underestimated at the CANDELS wavelengths; therefore we do not include this AGN when presenting results from SED fits in this paper.

\begin{figure}[ht!]
\includegraphics[width=0.43\textwidth]{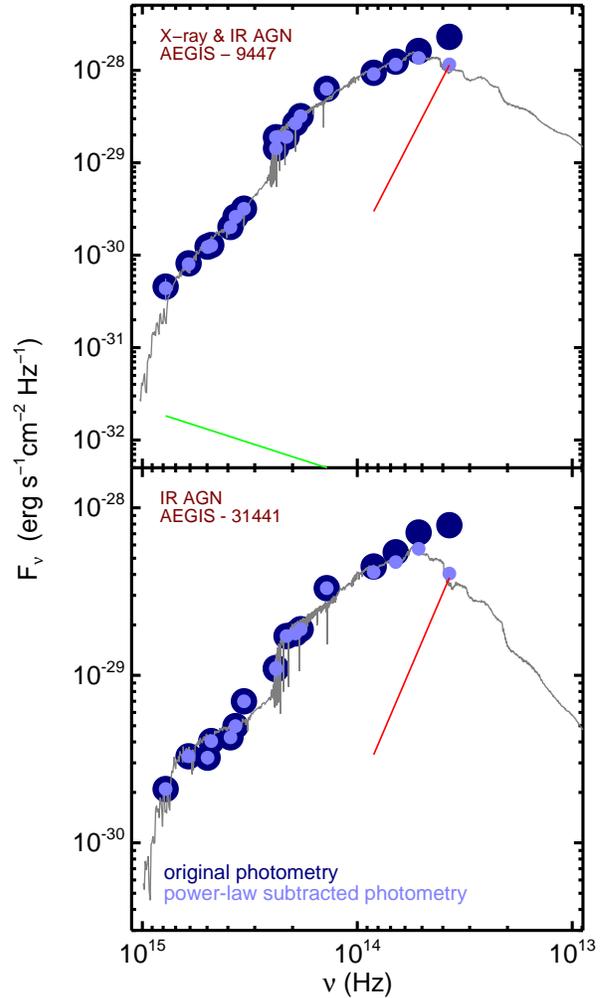}
\caption{The observed photometry (dark blue) and power-law subtracted photometry (light blue) for two AGN in the MOSDEF sample. The green line shows the power-law subtracted from the original photometry at rest-frame wavelengths $< $ 1 \um,  and the red line shows the power-law subtracted from the original photometry at rest-frame wavelengths $> $ 1 \um. The top panel shows an X-ray and IR AGN, where subtracting a blue power-law at UV and optical wavelengths and a red power-law at MIR wavelength results is a reduced $\chi^{2}$ smaller than the fit to the original photometry. The bottom panel shows an IR AGN where the best fit requires subtracting only a red power-law, as shown.}
\label{fig:sed}
\end{figure}

Light from the AGN may contribute to the observed SED, particularly at UV and IR wavelengths, which can impact our estimates of the SFR and stellar mass of the host galaxy. The Wien tail of the blackbody radiation from dust grains near the SMBH can be fit with a red power-law at mid-IR wavelengths. Therefore, to take into account any possible contribution from the AGN, we subtract power-laws with varying slopes and normalizations in both the UV and mid-IR from the observed photometry and re-run FAST on the remaining flux.  We then choose the fit with the lowest reduced $\chi^{2}$ as the best fit across all of the possible inputs (including no power-law subtraction, i.e., all galaxy light).

To create the grid of power-law SEDs to subtract, for both UV and mid-IR wavelengths, we allow the normalization to vary from 0 to 100 \% in terms of the observed flux in the U and IRAC channel 3 (5.8 \um) bands. We assume power-laws of the form  $F_{\rm\nu} \propto  \nu^{\alpha}$, where we allow ${\alpha}$ to range from 0 to 0.5 for the UV and -5 to -0.5 in the mid-IR.  We note that \cite{ivezi2002} considered a wider range of the optical spectral indices, $-2 \le \alpha \le 0.5$, based on the quasars in SDSS sample, but we consider a more limited range as the AGN in our sample are all lower luminosity and are type 2 AGN that are not expected to be dominated by the AGN component in the optical. We subtract the blue power-law from the photometry at rest-frame wavelengths $< $ 1 \um.

For subtracting the red power-law corresponding to AGN contribution at mid-IR wavelengths, we initially considered slopes within the range of -3 to -0.5, following \cite{donley2012}. 
However, we found that a redder slope was often needed to describe the observed slope of the SED for the two reddest IRAC channels (3 and 4) and thus adjusted our limits to allow for slopes as red as $\alpha_\mathrm{IR}>-5$.  We subtract the IR power-laws from the photometry at rest-frame wavelengths $> $ 1 \um \; to avoid any unnecessary subtraction from other bands. The power-law subtraction allows us to correct for any possible contamination from AGN in our SED-derived host galaxy properties.

Figure \ref{fig:sed} shows two examples of the original 3D-\textit{HST} photometry and the power-law subtracted photometry for an AGN identified at both X-ray and IR wavelengths (top panel) and an IR-only AGN (bottom panel). The green line shows the power-law subtracted from the original photometry at rest-frame wavelengths $< $ 1 \um, while the red line shows the power-law subtracted from the original photometry at rest-frame wavelengths $> $ 1 \um. In the top panel, subtracting the blue power-law at UV and optical wavelengths and the red power-law at MIR wavelengths results in a better fit. For the IR AGN in the bottom panel subtracting only a red power-law at MIR wavelengths results in a better fit.

Using this method leads to a more robust estimate of SFR and stellar mass for AGN host galaxies. \cite{santini2012enhanced} find that for type 2 AGN, stellar mass estimates derived from pure stellar templates are within a factor of two of the estimates derived including both stellar and AGN templates (with a mean difference of zero). Subtracting the AGN contribution from the original photometry as we do here leads to a 0.13 dex decrease in the average SFR and an 0.03 dex (in logarithmic space) decrease in the average stellar mass of the AGN in our sample. 

Here we run the FAST code without using the so-called template error function, which can be used to account for any wavelength-dependent mismatch between the observed photometry of sources and the templates used \citep{Brammer2008,kriek2009,kriek2015}. We find that the SFRs and stellar masses estimated from FAST without the template error function after subtracting power-laws are consistent with those derived from the original photometry using the template error function; this indicates our method for estimating SFR and stellar mass is robust.

\begin{figure}[t]
\includegraphics[width=0.47\textwidth]{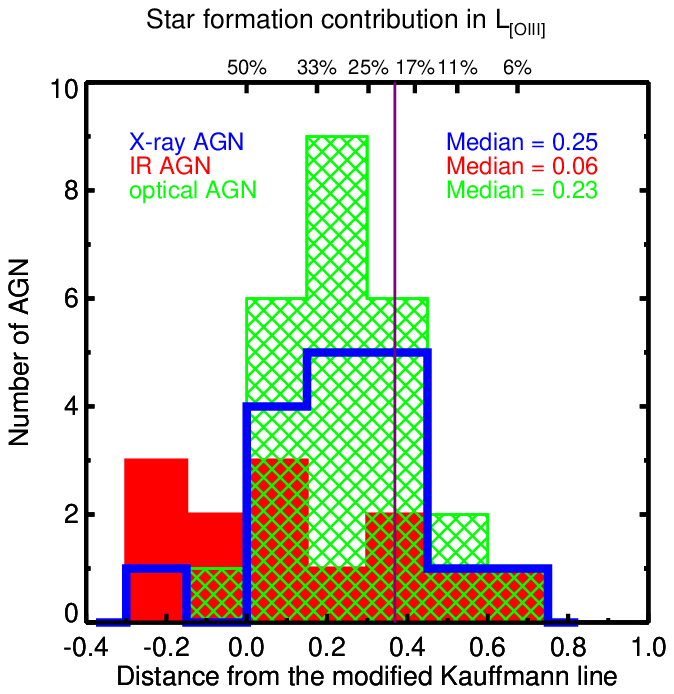} 
\caption{Histograms of the distance (in both \xbpt and \ybpt) of each AGN population in the BPT diagram from the modified version of \cite{kauffmann2003} line (see text for details). The top x axis indicates the fractional star formation contribution to \lo, according to \cite{kauffmann2009}. The median distance of our AGN sample is 0.17 dex 
which corresponds to a fractional contamination of $\sim$33\%. The purple line roughly shows the distance of the \cite{kewley2013} line, above which the contamination is  less than 20\%.}
\label{fig:distance}
\end{figure}

\subsection{\ox Luminosity as a Proxy of AGN Activity}
\label{sec:contamination}

The \ox emission line traces photoionized gas clouds in the narrow line region of AGN and is a good proxy for measuring nuclear activity 
\citep[e.g.][]{kauffmann2003,heckman2005}. However, star formation (SF) activity can also produce a narrow \ox emission line, which can contaminate the signal from the AGN. At low redshifts, studies have used different methods to attempt to disentangle AGN and SF contributions to the \ox emission line \citep[e.g.][]{kauffmann2009,wild2010,tanaka2011}. 
In this section we investigate whether these methods are applicable to higher redshift samples and estimate the contribution from SF to \ox luminosity in our sample.

\begin{figure*}
\centering
\includegraphics[width=0.8\textwidth]{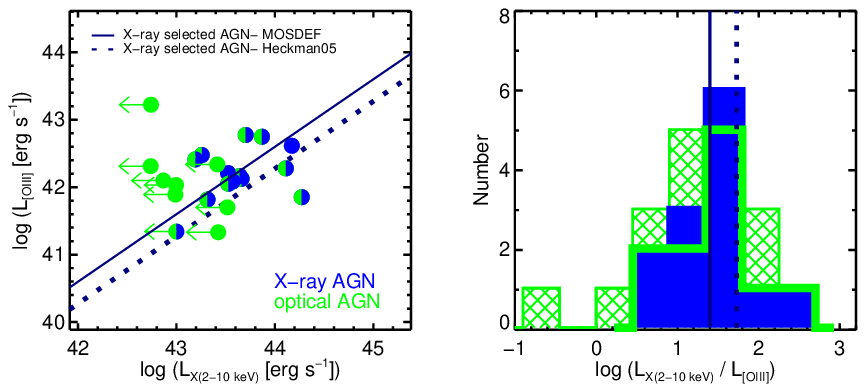}
\caption{\textit {Left:}\lo versus \lhx for X-ray (blue) and optical (green) AGN. The arrows shows an upper limit on \lhx for sources without X-ray detection. The solid light blue line shows the average \lratio (1.40 dex, with 0.44 dex standard deviation) for X-ray AGN in MOSDEF while the dotted blue line shows the ratio (1.73 dex, with  0.41 dex standard deviation) for a sample of X-ray selected AGN in \cite{heckman2005}. \textit{Right:} The distribution of \lratio for X-ray AGN and optically selected AGN in MOSDEF. The solid green histogram indicates \lratio for the optically selected AGN with X-ray identification, and the hashed histogram includes X-ray limits as well.}
\label{fig:lo_lx}
\label{fig:lo_lx}
\end{figure*}

\cite{kauffmann2009} and \cite{wild2010} use empirical techniques to estimate the contribution of SF to \lo by measuring the distance of each source in the BPT diagram from the main locus of star formation. \cite{kauffmann2009} find that the contribution to \lo from SF varies from $\sim$50\% for sources on the Kauffmann et al. line to $\sim$10\% for sources above the Kewley et al line. 

Alternatively, \cite{tanaka2011} use SED fitting to estimate the SFR for SDSS sources, and then convert this to an \ha luminosity using the \cite{kennicutt1998star} relation, with an additional power-law as an extinction factor estimated at the \ha wavelength. They then fit a relationship to the observed ratio of [{\ion{O}{3}}]/{{H$\alpha$}\;} with stellar mass and estimate \lo from star formation for individual sources.

However, in our sample at $z\sim2$, estimating the SF contribution to \ox is more challenging, as we do not always have high S/N spectroscopic measurements of all of the relevant emission lines for all of our AGN. Initially, we investigated the level of SF contamination for AGN in the BPT diagram, using the method of \citet{kauffmann2009}. Out of the 55 AGN in our sample, 31 have sufficient S/N to place them in the BPT diagram, though one of the two line ratios may be a limit.
For these sources, we measure the distance of each AGN from the Kauffmann et al. 
line. However, we first shift the Kauffmann et al. line $\sim0.1$ dex higher in log(\lybpt) to account for the overall offset of the MOSDEF galaxy sample compared to SF galaxies in SDSS (see Figure \ref{fig:bpt} and Section \ref{sec:optagn} above); though we note that this shift could be along log(\lxbpt)  as well \citep[see][]{shapley2015,Sanders2016}.
Figure \ref{fig:distance} shows the histogram of the distance of each AGN on the BPT diagram in our sample from this modified Kauffmann et al. line.

For the AGN that we can show on the BPT diagram, we find a  median distance of 0.17 dex  from the modified Kauffmann et al. line. This median point (log(\lxbpt) = -0.39, log(\lybpt) = 0.36) lies in the Seyfert region of the BPT diagram, so we use the relevant trajectory from \cite{kauffmann2009} to find the contribution of SF to the \ox emission. The top x axis in Figure \ref{fig:distance} shows the fractional star formation contribution to \lo. The median contribution for our AGN is $\sim$33\%. 
The purple line in Figure \ref{fig:distance} roughly shows the distance of the \cite{kewley2013} line. This indicates that AGN above the \cite{kewley2013} line should have a fractional SF contamination of less than 20\%.

We also estimate the median SF contamination to \lo using the method of \cite {tanaka2011}, as described above and find a median contamination of 30\%. However, using this method we find a prohibitively large scatter which unfortunately indicates that we cannot use this method to apply a correction for SF on a source by source basis.

Here, we use \lo as a proxy of AGN activity as the majority of the \ox flux is from the AGN (both methods described above estimate a $\sim32\%$ contribution from SF activity).  Additionally, almost all of the AGN in our sample lie in the AGN region of the BPT diagram (for optical AGN this is by definition), which indicates their emission lines are dominated by AGN radiation rather than SF. We do not correct \lo for SF contamination here, as we cannot apply the \cite{kauffmann2009} method since it requires reliable detections for each emission line, and using the \cite {tanaka2011} method results in a substantial additional scatter. As we discuss below in Section \ref{sec:results}, any trends that we find with \lo are also confirmed with \lx in our X-ray detected sample, such that none of our results should be substantially impacted by SF contamination to \ox.

\section{Results}
\label{sec:results}

In this section we consider AGN identification at different wavelengths and 
investigate AGN selection biases and host galaxy properties for different 
identification techniques. We first compare the luminosities of AGN selected at X-ray versus optical 
wavelengths. We then compare AGN host galaxy properties, such as SFR, 
dust content and stellar age, to a sample of inactive galaxies with the same
distribution of stellar mass as the AGN host galaxies. Finally, we investigate 
the relationship between AGN activity and SFR in individual AGN host galaxies at $z\sim2$. We emphasize that for any analysis with \ox measurements we restrict our sample to sources with $>3\sigma$ \redox detections, which includes 34 AGN and 374 galaxies.

 %%%%%%%%%%%%%%%%%%%%%%%%%%%%%%%%%%%%%%%%%%%%%%%%%%%%%%%%%%%%%%%%%%%%
 
\subsection{The Relationship between X-ray and Optical Emission} 
\label{sec:lx_loiii}

In this section, we address whether we can recover AGN that are absorbed at X-ray wavelengths by comparing \lx and \lo measurements for our samples of X-ray and optical AGN.

The \redox line, as one of the strongest narrow optical emission lines, provides a robust tracer of AGN power and (at least at lower redshifts) is less contaminated by emission due to star formation activity than the \ha line \citep[e.g.][]{heckman2004,Brinchmann2004}. The hard band X-ray emission is also a robust estimator of AGN power and can penetrate regions with low to moderate hydrogen column densities. However, X-ray emission will be suppressed in Compton thick or highly obscured AGN. Low luminosity AGN may also be missed due to the flux limit of the X-ray data especially towards the edges of the X-ray pointings.

In Figure \ref{fig:lo_lx} (left panel) we show 
\lo versus \lhx for X-ray (blue circles) and optically selected (green circles)
AGN in MOSDEF. For the optical AGN that do not have X-ray detections in the hard band, we use the upper limits on \lhx (at 95\% confidence level).
Here we consider only AGN with significant \redox detections.\footnote{The \ox line for sources not shown in this figure is not necessarily low flux; in many cases it is simply impacted by a sky line. We also note that even if we do not detect \ox for a given source, we can still in some cases optically identify it as an AGN, if it has a high [NII]/H$\alpha$ ratio.}
The blue solid line shows the average \lratio for X-ray AGN in MOSDEF. The dotted line shows the measurement of \lratio for an X-ray selected sample at $z<0.2$ by \cite{heckman2005}.

We note that the AGN \redox luminosity in our sample is corrected for dust reddening. To determine the correction factor, we calculate the color 
excess from the Balmer decrement and combine this with the value of the MOSDEF dust attenuation curve at 5008 \AA \citep{reddy2015}. This correction results in an average increase of $\sim$ 0.17 dex in \ox luminosity for the AGN in our sample.  The \ox luminosities from  \cite{heckman2005} are not corrected
for dust reddening. We also note that the X-ray selected AGN in the \cite{heckman2005} sample are detected in the 3--20~keV hard X-ray band, and we converted their luminosity to \lhx assuming $\Gamma$ =1.9. 

\begin{figure*}
\includegraphics[width=0.96\textwidth]{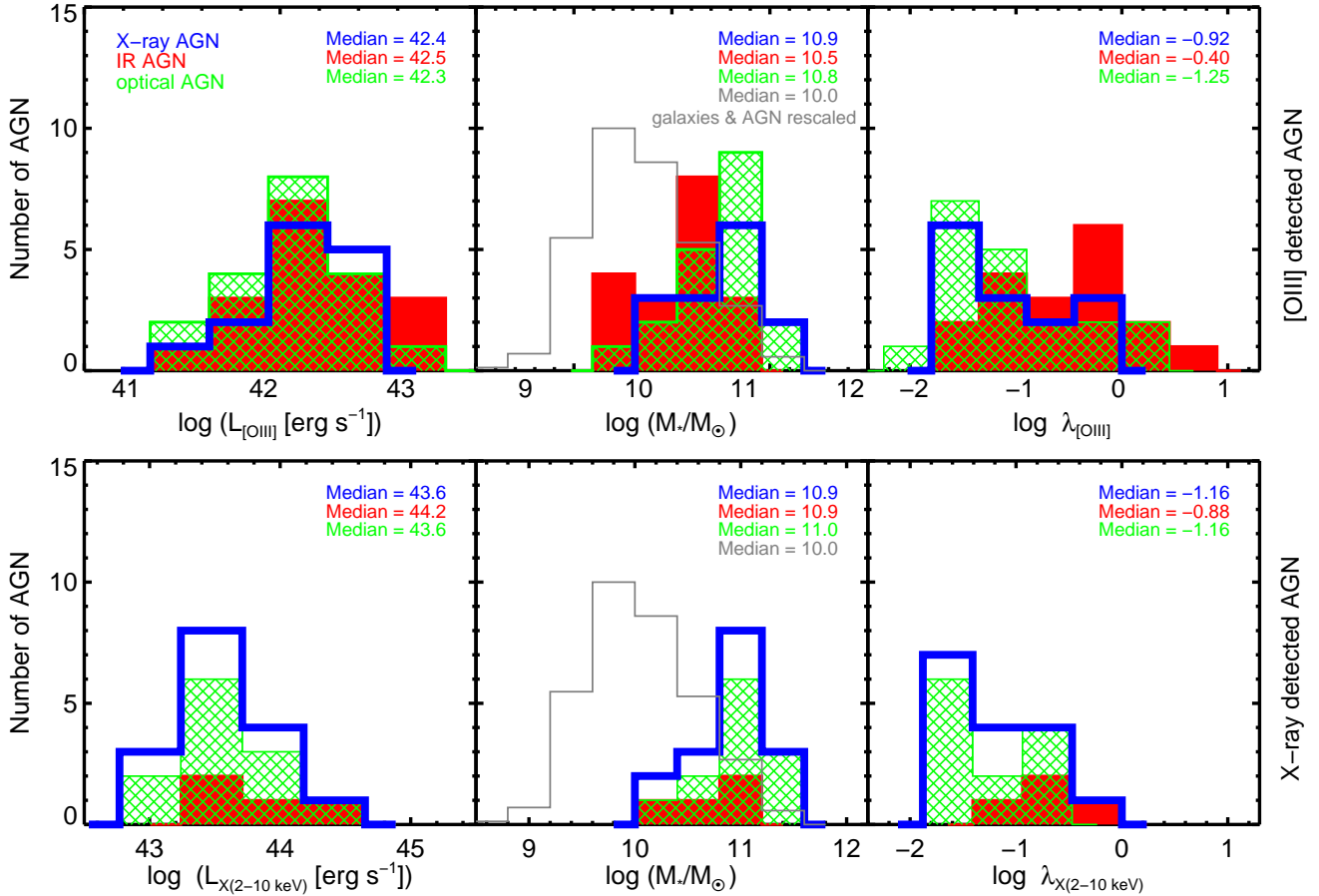}
\caption{The observed luminosity (\textit{left}), stellar mass (\textit{center}), and specific accretion rates (\textit{right}) distributions of X-ray (blue), IR (red), and optical (green) AGN with significant \ox detection in top row and significant hard X-ray detection in the bottom row. The median values are given in each panel.
AGN selected at all three wavelengths in 
our sample have very similar \lo  (and \lx) distributions (left panels). In the middle panels the gray histograms indicate the stellar mass distributions of our entire sample (galaxies and AGN). The stellar mass distributions clearly reflect the bias against AGN identification in low mass galaxies.}
\label{fig:LOIII_hist}
\end{figure*}

In the right panel of Figure \ref{fig:lo_lx} we show the distributions of \lratio for X-ray and optical AGN samples. For the X-ray selected AGN in MOSDEF, the average \lratio is 1.40 dex (with a standard deviation of 0.44 dex). This measurement is consistent with the average ratio in \cite{heckman2005} (1.73 dex with a standard deviation of 0.41 dex), considering the uncertainties and the lack of the reddening correction in \cite{heckman2005}. \cite{Trouille2010} with a larger sample of X-ray selected AGN at $z<0.85$ also find a similar average \lratio (1.46 dex, with 0.6 dex standard deviation), and indicate that the fraction of X-ray AGN that are not identified in optical diagnostics varies between 20-50\% depending on the line used for classifying optical AGN (\citealt{kauffmann2003} versus  \citealt{Kewley2001}). \cite{Trouille2010} argue that this misidentification could be due to the complexity of the structure of the narrow-line region, which could result in escape of many of the ionizing photons, and therefore lower \lo in some sources.

Considering a sample of optically selected AGN,\footnote{We note that the optical AGN in \cite{heckman2005} are selected based on an \redox flux limit rather than the BPT diagram.} \cite{heckman2005} find a large scatter in \lratio and identify a population of type 2 AGN that are bright at \redox but under-luminous in X-ray. As shown in Figure \ref{fig:lo_lx}, in MOSDEF we only identify two AGN with low \lratio indicating the X-ray emission is heavily obscured.

The bulk of optically selected AGN in our sample have a similar \lratio to the X-ray selected sample. The majority of optically selected AGN with limits on \lx still follow the overall \lratio trend  and are consistent with being intrinsically lower luminosity AGN. Thus at $z\sim2$ the optical selection method is effective at identifying lower luminosity AGN that may be missed by X-ray surveys due to the limited (and variable) depth of the X-ray data. However, deeper X-ray data could reveal that these sources are under-luminous at X-ray wavelengths and thus are candidate obscured AGN.

Overall, we find that the relationship between \lx and \lo in our sample at  $z\sim2$ is consistent with that of the \cite{heckman2005} local X-ray AGN sample.
However, only $\sim$50\% of our optically selected AGN are detected at X-ray wavelengths (see also Figure \ref{fig:venn}). 
In part, this is due to the fact that the X-ray data does not have uniform depth across our fields, unlike the more uniform \ox sensitivity in the MOSDEF spectra. 
Thus, at these redshifts, optical selection may be more effective at identifying AGN, especially lower luminosity sources, but does not obviously identify significant populations of heavily absorbed AGN.
We also find that $\sim$75\% of the X-ray AGN are selected at optical wavelength, indicating that optical selection is relatively complete but can miss some AGN that are found by X-ray surveys. This is mainly due to contamination of the optical spectra at $z\sim2$ from the sky lines, as we discuss below in Section \ref{sec:depth}.

\begin{figure*} [t]
\centering
\includegraphics[width=0.87\textwidth]{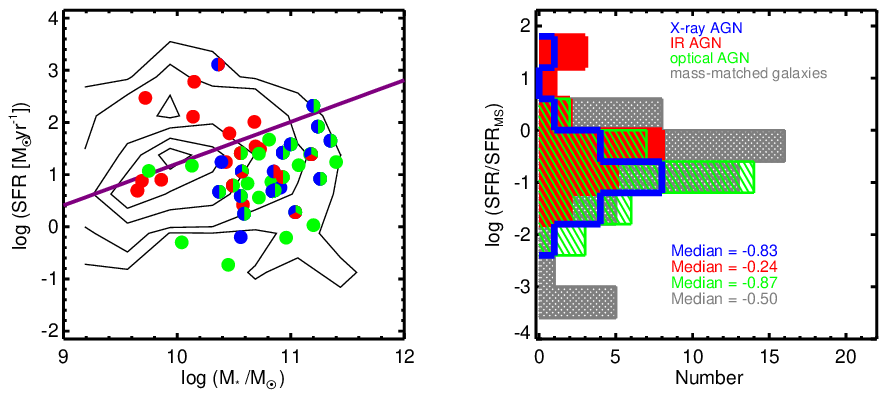}
\caption{{\it Left:} SFR versus stellar mass for the full MOSDEF galaxy sample (contours) and X-ray (blue), IR (red), and optically-selected (green) AGN host galaxies. The purple line shows the relation from \cite{shivaei2015} for the main sequence of star formation of MOSDEF galaxies at $z<2.6$. Compared to the full galaxy sample, AGN host galaxies span a similar range of SFR but a more limited range of stellar mass. {\it Right:} The \sfratio distributions in a stellar mass-matched inactive galaxy sample (gray) compared to the various AGN samples. The median \sfratio are given for each population. IR AGN appear to be biased towards higher  \sfratio. Based on the KS test, we find that the IR AGN have a different distribution of SFR/$\mathrm{SFR_{MS}}$ (at the $>2\sigma$ equivalent confidence level) compared to either the optical AGN ($p=0.004$) or the X-ray AGN ($p=0.02$) samples.
}
\label{fig:sfr_mass}

\end{figure*}

%%%%%%%%%%%%%%%%%%%%%%%%%%%%%%%%%%%%%%%%%%%%%%%%%%%%%%%%%%%%%%%%%%%%%%%%%%%%%%%%%%%%%%%%%%%%%%%%%%%%%%%%%%%%%%%%%%%%%%%%%%%%%%%%%%%%%%%%%%%%

\subsection{AGN Luminosities and Specific Accretion Rates}
\label{sec:oiii_hist}

To further investigate the differences between AGN samples identified at different wavelengths at $z\sim2$, 
in Figure \ref{fig:LOIII_hist} we compare the distributions of AGN luminosities (left panels), host stellar masses (central panels), and specific accretion rates (right panels) for our samples of X-ray AGN (blue), IR AGN (red) and optical AGN (green). In the middle panels, we additionally show the rescaled histogram of the stellar mass distribution of our entire galaxy and AGN sample that includes 531 sources in gray.
In the top row of Figure \ref{fig:LOIII_hist}, we show only those sources with a significant \ox flux in the MOSDEF data (and thus measured \lo), which results in a sample of 34 sources.
In the bottom row of Figure \ref{fig:LOIII_hist}, we show only those sources with a significant hard X-ray detection (and thus measured \lx), resulting in a sample of 16 sources (after excluding broad-line AGN and sources with soft band detections only). As noted above, a single source can be identified as an AGN at X-ray, IR and optical wavelengths, therefore the same object can be included in multiple distributions here. 
By construction, all sources in the bottom panels are identified as X-ray AGN, whereas in the top panels we include sources identified as X-ray and IR AGN where we are able to measure \lo\ but do not identify the source as an AGN based on our BPT diagnostics.

The specific accretion rate (shown in the right panels of Figure \ref{fig:LOIII_hist}) traces the rate of SMBH growth relative to the stellar mass of the host galaxy, providing an indicator of how rapidly a galaxy is growing its black hole \citep[see][]{aird2012primus}. By calculating specific accretion rates, we can account for any differences in the stellar masses of the host galaxies of AGN selected at different wavelengths, revealing  differences in the types of AGN that are selected with each method that may not be apparent from the observed luminosities.

The specific accretion rate is calculated from either the \lo or \lx and is given by
\begin{equation}
\lambda_\mathrm{band}=\frac{k_\mathrm{band} L_\mathrm{band}}{1.3\times 10^{38}\times 0.002 \dfrac{M_*}{M_\odot} }
\end{equation}

where $k_\mathrm{band}$ is the corresponding bolometric correction. We adopt a single bolometric correction at each wavelength. At optical wavelengths we use a mean bolometric correction of 600 from \cite{kauffmann2009}, which corresponds to the mean correction for extinction-corrected \redox luminosity \citep[see also][]{Netzer2009,LaMassa2010}. For sources with 
X-ray detections we use a constant bolometric correction of $k_\mathrm{X(2-10\; keV)}=25$. We also estimated the bolometric luminosity using the luminosity-dependent bolometric corrections from \cite{Hopkins2007} and \cite{Lusso2012} for type 2 AGN, but this does not alter the overall trends seen in Figure \ref{fig:LOIII_hist} when using a single bolometric correction. 
The denominator in the above equation is chosen such that the units of $\lambda_\mathrm{band}$ approximately correspond to the Eddington ratio (assuming a single scaling between SMBH mass and total stellar mass). 

For AGN with significant \ox detections, the median statistical uncertainties on  \lo, stellar mass, and \lamo are 0.04 dex, 0.10 dex, and 0.12 dex, respectively.  For AGN with X-ray detections, the median uncertainties on \lx, stellar mass, and \lamxx are 0.20 dex, 0.07 dex, and 0.23 dex.  However, there may be additional uncertainties in \lamo and \lamxx from the bolometric corrections (which may depend on Eddington ratio, see \citealt{Vasudevan2007}) that could result in larger uncertainties.

We use a Kolmogorov--Smirnov (KS) test to compare the distributions shown in Figure \ref{fig:LOIII_hist} for the different AGN selections and assess if there are significant differences (requiring a $p$-value $<0.05$, corresponding to a $>2\sigma$ equivalent confidence level). 
Based on our KS tests, we find no evidence for significant differences in the distributions between any two samples, indicating that the distributions of luminosities, host stellar masses and specific accretion rates for X-ray, IR, and optical AGN are all statistically consistent with being drawn from the same parent population. 

However, the lack of significant differences could be due to our relatively small sample sizes.
In Figure \ref{fig:LOIII_hist} there are indications of differences in these distributions, probed here for the first time at $z\sim2$, that appear consistent with previously identified selection biases at lower redshifts \citep[e.g. $z\sim0.1-1$:][]{mendez2013primus}.
In general, there is a bias against AGN identification in relatively low mass galaxies. 
For IR selection there may be an additional bias against the most massive galaxies in our sample. 
IR selection appears to identify AGN with, on average, lower stellar masses and higher specific accretion rates i.e., sources where the light from the AGN dominates over the host galaxy.

X-ray selection is able to probe low specific accretion rates which may introduce a bias toward higher stellar mass host galaxies \citep[e.g.][due to the X-ray flux limits, low specific accretion rate sources will not be identified in lower mass galaxies]{aird2012primus}.
Optical selection---a key additional probe with our MOSDEF sample---appears to identify AGN with similar properties to the X-ray selected population i.e., down to low specific accretion rates and generally in higher stellar mass galaxies.

Overall, our results show that with optical diagnostics we are able to identify less powerful, low accretion rate AGN that may not be identified at other wavelengths (either due to obscuration or limited sensitivity).
We also find that IR AGN selection preferentially identifies powerful AGN that are hosted in relatively lower mass galaxies, compared to optical and X-ray AGN selection. Each of these identification methods at different wavelengths are incomplete and suffer from selection biases, therefore a combination of identification methods can provide a more complete picture of AGN properties.

 %%%%%%%%%%%%%%%%%%%%%%%%%%%%%%%%%%%%%%%%%%%%%%%%%%%%%%%%%%%%%%%%%%%%%%%%%%%%%%%%%%%%%%%%%%%%%%%%%%%%%%%%%%%%%%%%%%%%%%%%%%%%%%%%%%%%%%%%%%%%

\subsection{AGN Host Galaxy Properties}
\label{sec:host}

In this section we investigate the properties of the host galaxies of our AGN samples in more detail and compare with a sample of inactive galaxies from MOSDEF. 

The MOSDEF galaxy sample spans a wide range in both SFR $(-1\lesssim \log (\frac{\mathrm{SFR}}{\mathcal{M}_\odot \; \mathrm{yr}^{-1}}) \lesssim 3)$ and stellar mass ($8\lesssim \log(\frac{\mathcal{M_{*}}}{{\mathcal{M}_\odot}})\lesssim 12$). From the data in the first two years of the MOSDEF survey, we identify 481 galaxies with an average stellar mass of  $\sim10^{10} \; \mathcal{M}_\odot$. While the AGN host galaxies span the full range of SFRs of the galaxy sample, they span a more limited range in stellar mass (less than 2 orders of magnitude) with an average stellar mass of $10^{10.6}\; \mathcal{M}_\odot$ for their host galaxies. The observational bias against AGN identification (at any wavelength) in low mass galaxies restricts our AGN sample to relatively massive host galaxies \citep[e.g.][]{aird2012primus}.

Since stellar mass correlates with other physical properties such as metallicity, age, and SFR of the galaxy, we construct a stellar mass-matched control sample of inactive galaxies for comparative analysis with the AGN host galaxies. To create this sample, we bin the AGN host galaxies in narrow intervals of $\Delta \log(\frac{{\mathcal{M_{*}}}}{{\mathcal{M}_\odot}})$ = 0.05 dex and select 50 inactive galaxies from the full galaxy sample to create a sample with the same stellar mass distribution as the AGN host galaxies.

In the left panel of Figure \ref{fig:sfr_mass}, we show the SFR versus stellar mass distributions of AGN (colored points) in MOSDEF compared to the full galaxy sample (contours). There is a well known, positive correlation between the SFR and stellar mass of galaxies, known as the ``main sequence'' of star formation \citep[e.g.,][]{elbaz2007,noeske2007,karim2011star,Whitaker2012,shivaei2015}. The purple line in this figure shows the relation for the main sequence of star formation, based on SED fitting, for MOSDEF galaxies at $1.4<z<2.6$ from \cite{shivaei2015}:
\begin{equation}
\log (\frac{\mathrm{SFR}}{\mathcal{M}_\odot \; \mathrm{yr}^{-1}}) = (0.80 \pm 0.05) \times 
\log (\frac{\mathcal{M_{*}}}{\mathcal{M}_\odot}) - (6.79 \pm 0.55)
\label{eq:MS}
\end{equation}

\begin{figure} [t]
\includegraphics[width=0.45\textwidth]{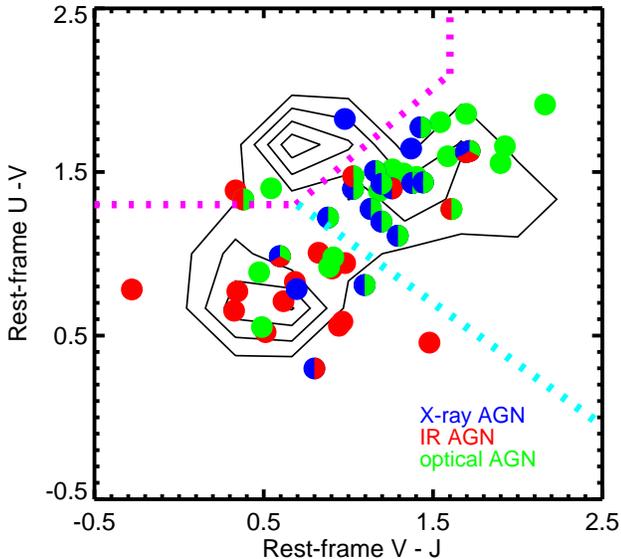}
\caption{The rest-frame U-V versus V-J color diagram, where contours shows mass-matched inactive galaxies and the blue, red, and green circles show X-ray, IR, and optical AGN. The dotted magenta line isolates quiescent galaxies using criteria from \cite{williams2009}, while the dotted cyan line shows the demarcation from \cite{kriek2015} for dividing dusty versus 
less dusty star-forming galaxies. Considering the errors in Table \ref{tab:fraction}, the fraction of AGN in each part is not significantly different than their mass-matched galaxies.
Comparing the fraction of IR and optical AGN indicates that a higher fraction of IR AGN are in less dusty star-forming region (at $3.8\sigma$ significance) and a higher fraction of optical AGN are in dusty star-forming region (at $ 4.2\sigma$ significance).}

\label{fig:uvj}
\end{figure}

Figure \ref{fig:sfr_mass} shows that AGN exist in galaxies over the full range of SFR probed by the MOSDEF sample. We find that the majority of optical and X-ray AGN host galaxies lie below the main sequence of star formation, while IR AGN host galaxies are found both above and below the main sequence. 

In the right panel of Figure \ref{fig:sfr_mass}, we show the distribution of \sfratio which is the relative offset of the SFR from the main sequence at the stellar mass of the host galaxy. We additionally show the \sfratio distribution for our mass-matched inactive galaxy sample. The median \sfratio for each sample is given in the figure. The median statistical uncertainty on $\log$(SFR/SFR${_\mathrm{MS}}$) for AGN  is 0.23 dex and 0.34 dex for the mass-matched galaxies. A KS test shows that the \sfratio distribution of the AGN host galaxies is not significantly different from our stellar mass-matched inactive galaxy sample. We further consider three control samples of inactive galaxies, each mass-matched with AGN identified at each wavelength, and compare the physical properties of individual AGN population with their mass-matched galaxies. We find that the distribution of \sfratio for each AGN sample is consistent with their inactive mass-matched galaxies.
Comparing between the various AGN samples (using KS tests), we find that the distributions of \sfratio of X-ray and optical AGN are statistically consistent with each other. However, we find the distribution of \sfratio for IR AGN is different at (at the $>2\sigma$ equivalent confidence level) compared to the optical AGN ($p=0.004$) or the X-ray AGN ($p=0.02$) samples.

To further investigate the host galaxy properties, we consider the rest-frame U--V versus V--J color 
diagram (UVJ color). This diagram is commonly used to distinguish quiescent galaxies from star-forming
galaxies with different dust content \citep[e.g][]{williams2009,whitaker2011,muzzin2013}. To estimate rest-frame colors we use the EAzY code \citep{Brammer2008} by interpolating between the observed photometric bands \citep[see][]{kriek2015}, with the AGN contributions subtracted. In Figure \ref{fig:uvj} we show our AGN (colored points) in UVJ space, along with mass-matched MOSDEF galaxies (contours). The dotted magenta line in this figure shows the region isolating the quiescent galaxies, identified using criteria from \cite{williams2009}:

\begin{align}
  &U-V > 1.3 \\ 
  &U-V = 0.88 \times(V-J)+0.69 \\
  &V-J < 1.6
   \label{q_uvj} 
\end{align}

The cyan dotted line shows the demarcation from \cite{kriek2015} that divides star-forming galaxies into those that are red and dusty from those that are blue and less dusty. 

Table \ref{tab:fraction} indicates the fraction of AGN in each region of UVJ space, compared to the fractions of our mass-matched galaxy sample that fall in each region. The errors on the fractions are estimated from bootstrap resampling. Given the errors, there is not a significant difference between the fractions of AGN and mass-matched galaxies in different regions of UVJ space.

\begin{table}
\caption{ The fraction of AGN and, stellar mass-matched galaxies in different parts of UVJ space}
\begin{center}
\renewcommand{\arraystretch}{2}
\begin{tabular}{m{1.in} m{1.in} m {0.8in}}
\hline \hline
Region & AGN & mass-matched \;\;galaxies   \\
\hline \hline 
quiescent & 8\% $\pm$ 4\% & 14\% $\pm$ 4\% \\  
  less dusty SF & 39\% $\pm$ 7\% & 49\% $\pm$ 6\% \\
  dusty SF & 53\%  $\pm$ 7\% &  37\% $\pm$ 6\% \\    \hline
 
 \end{tabular}
\end{center}
\label{tab:fraction}
\end{table}

The location of galaxies in UVJ space depends sensitively on stellar mass; in particular within the star-forming population, dusty galaxies are more massive \citep[e.g][]{Williams2010}. Here, we find that AGN have a similar distribution in UVJ space to a stellar mass-matched inactive galaxy sample. Figure \ref{fig:uvj} also shows that the majority of optical AGN (73\%) and X-ray AGN (72\%) are identified in dusty star-forming galaxies, while the majority of the IR AGN are identified in less dusty star-forming galaxies (68\%).

Considering the sensitivity of the UVJ diagram to stellar mass, and the fact that each AGN population has a different mass distribution, we compare the distribution of UVJ colors for each of our three AGN samples to an appropriately mass-matched galaxy sample. For the X-ray and IR AGN samples we find that their distribution in UVJ space is consistent with their corresponding mass-matched galaxy sample, indicating that the higher density of IR-AGN in the non-dusty star-forming region in Figure \ref{fig:uvj} can be attributed to the lower stellar masses (on average) of the hosts of IR-selected AGN. The fraction of optical AGN in the dusty star-forming region is higher than their mass matched galaxies (at $2.9\sigma$ significance), which we discuss below in Section \ref{sec:discussion}.

%%%%%%%%%%%%%%%%%%%%%%%%%%%%%%%%%%%%%%%%%%%%%%%%%%%%%%%%%%%%%%%%%

\begin{figure*} [t]
\includegraphics[width=0.95\textwidth]{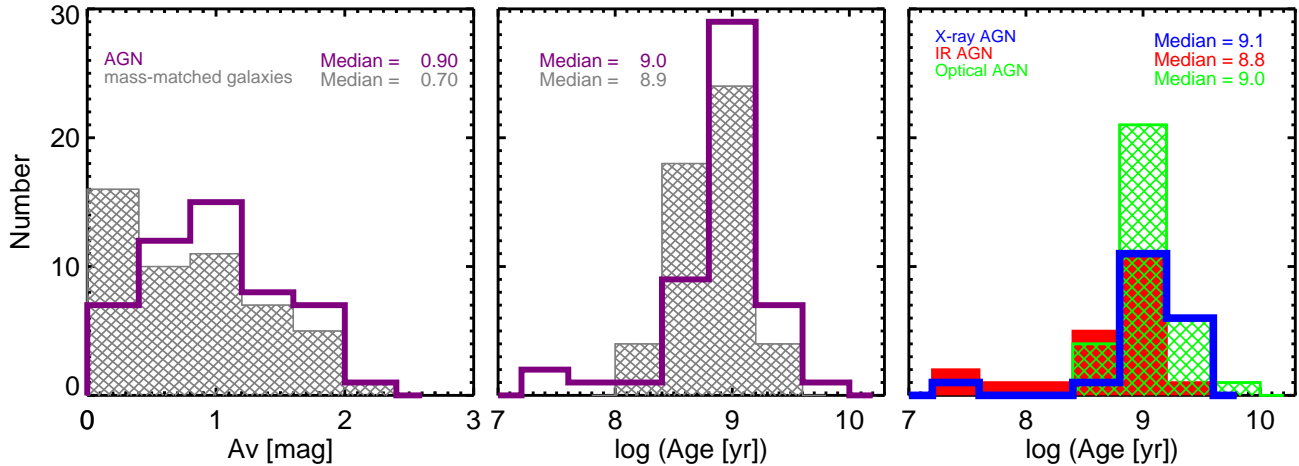}
\caption{{\it{Left:}} The dust extinction (A$_v$) distributions; {\it{Middle:}} The stellar age distribution for full MOSDEF AGN (purple) and mass-matched galaxy samples (grey). The KS test shows that the AGN and mass-matched galaxy samples have statistically similar A$_v$ and age distributions. The median values with the standard errors are given for each population. {\it{Right:}}The stellar age distribution in each AGN population. IR AGN reside in galaxies with younger stellar population compared to the optical AGN host galaxies at $> 2\sigma$ significance ($p=0.02$).}
\label{fig:av_age}
\end{figure*}

In Figure \ref{fig:av_age} we compare the dust extinction and stellar age derived from SED fitting (after subtracting the AGN contribution) for AGN host galaxies with the mass-matched inactive galaxies. We illustrate the distributions of visual extinction ($A_V$) (left panel) and stellar age (middle panel) for AGN and mass-matched galaxies. Additionally, we show the distributions of stellar age for individual AGN populations in the right panel. The median values of each distribution are given in the figure. The median statistical error on $A_V$ is 0.40 magnitudes for the AGN and 0.30 magnitudes for the mass-matched galaxies.  The median error on $\log$(stellar age) is 0.30 dex for the AGN and 0.13 dex for the mass-matched galaxies.

\begin{figure*} [t]
\centering
\includegraphics[width=0.8\textwidth]{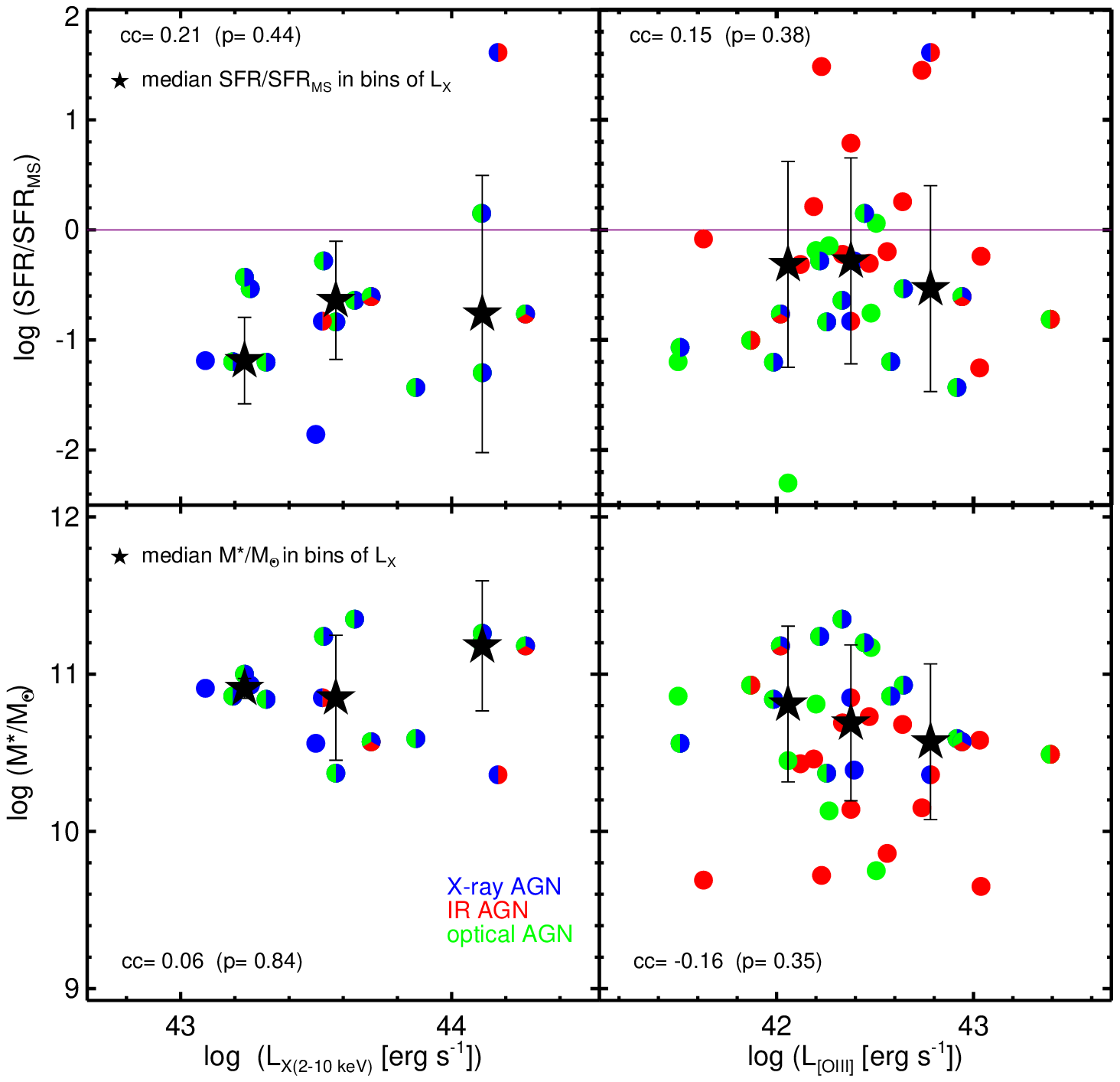}
\caption{Left: The AGN host galaxy SFR relative to the main sequence of SFR (\sfratio, top) and 
host galaxy stellar mass (bottom)  as a function of \lx (left) and \lo (right, for AGN with significant \ox detections).
The X-ray, IR and optical AGN are shown respectively with blue, red and green circles; the black stars indicate the median \sfratio and stellar mass in bins of \lx (or \lo), with the error bars showing the standard deviation on the median values. The purple horizontal line in the upper panels shows
the main sequence of star formation, based on SED fitting, for MOSDEF galaxies at $1.4<z<2.6$ from \cite{shivaei2015}, used here to define a \sfratio of zero. The correlation coefficients are given in each panel.}
\label{fig:sfr_lagn}
\end{figure*}

 The full (non mass-matched) MOSDEF galaxy sample has a median $A_V$=0.5 magnitude and a median stellar age of 10$^{8.6}$ yr, with distributions in $A_V$ and stellar age that are significantly different from 
the AGN sample. The median dust extinction and age are very similar in AGN and mass-matched inactive galaxies populations, and KS tests show that the two populations do not have statistically different distributions in either parameter. The similarity of these distributions is not entirely unexpected, given that there are strong correlations between stellar mass and both extinction and stellar age.

We note that stellar age estimation is sensitive to various parameters in SED fitting, in particular, the star formation history models. However, the distribution of $\tau$ (the characteristic star formation timescale) for the MOSDEF AGN is not significantly different from the distribution of $\tau$ for the mass-matched galaxies.  We also find a similar {\it joint} distribution in stellar age and dust extinction between the AGN and mass-matched galaxies, in that galaxies with younger stellar populations are dustier than galaxies with older stellar populations.

In the right panel of Figure \ref{fig:av_age} we show stellar age distributions for the separate AGN populations in our sample and find a median stellar age of $\sim$ 1.3 Gyr, 630 Myr and 1 Gyr, respectively, for X-ray, IR and optical AGN. KS tests show a $2\sigma$ ($p$=0.02) significance level difference in their age distributions that  is most likely due to the stellar mass selection biases, with IR AGN being identified in relatively lower mass galaxies, compared to optical and X-ray AGN.

Overall, our results indicate that the distributions of SFR, dust, and stellar age in AGN and mass-matched inactive galaxies are very similar, and that the key parameter in finding these similar distributions is the stellar mass of both populations. Stellar mass also plays an important role in AGN identification. While IR AGN are biased against the most massive galaxies, we can identify them in less dusty galaxies with younger stellar population and relatively high star formation activity. In contrast, optical AGN are identified in dusty massive galaxies with older stellar populations and lower star formation activity.

%%%%%%%%%%%%%%%%%%%%%%%%%%%%%%%%%%%%%%%%%%%%%%%%%%%%%%%%%%%%%%%%%%%%%%%%%%%%%%%%%%%%%%%%%%%%%%%%

 \subsection{The Relationship between Star Formation and AGN Activity}
\label{sec:correlation}

We now investigate whether there is a connection between star formation activity and AGN activity for individual sources in our $z\sim2$ sample. To trace the AGN activity we use \lx for those AGN with X-ray detections and \lo for AGN with significant \ox measurements. 
To quantify the relation between AGN luminosity and SFR, we calculate the correlation coefficient and the corresponding significance using the $r-correlate$ routine in IDL, which computes the Spearman's rank correlation coefficient (cc) and the significance of its deviation from zero ($p$).

As noted in Section \ref{sec:host}, the SFRs and stellar masses of galaxies are known to be correlated \citep[e.g.,][]{elbaz2007,karim2011star,shivaei2015}. Therefore an underlying correlation between AGN luminosity and host galaxy stellar mass, if it existed, could result in a correlation between SFR and AGN luminosity.
To take this stellar mass-dependent effect into account, instead of quantifying any correlation between SFR and AGN luminosity, we use \sfratio (the relative offset of the SFR from the main sequence at the stellar mass of the host galaxy). We use equation \ref{eq:MS}, which defines the star-forming main sequence for the MOSDEF sample for galaxies at $1.4<z<2.6$ from \cite{shivaei2015}. 

In Figure \ref{fig:sfr_lagn}, we show \sfratio versus \lx for X-ray AGN host galaxies in the top left panel and stellar mass versus \lx in the lower left panel. The right panels show 
\sfratio versus \lo (top) and stellar mass versus \lo (bottom) for AGN with 3$\sigma$ \ox detections. We also show the median \sfratio and stellar mass in bins of \lx and \lo with the black stars, where the error bars indicate the standard deviation of the median in each luminosity bin. 
As shown in the figure, we find no significant correlation between \sfratio and either \lx or \lo. 

We note again that we have not accounted for any contribution to \lo from star formation in our sample, as we are unable to correct for this on a source by source basis. In general, there is a positive correlation between SFR and \ox emission in the galaxy population \citep[e.g.][]{Mehta2015}. Although we do not find a significant correlation between \sfratio and \lo here, the possible contribution from star formation to \lo could produce a correlation and should thus be considered in any future studies with larger samples.

Overall, we do not find any significant correlations between SFR and AGN luminosity in our sample. However, a common gas supply for triggering and fueling both of these phenomena could play an important role in galaxy and AGN growth. With a larger sample (and the possibility of correcting \lo for star formation contributions), the connection between the growth of SMBHs and their host galaxies can be investigated more accurately.

\section{Discussion}
\label{sec:discussion}

In this paper we use data from the first two years of the MOSDEF survey, which includes 55 AGN identified with X-ray, IR, and/or optical diagnostics 
at $z\sim2$.

We investigate the selection of these AGN and their host 
galaxies properties. Below we first discuss the uniqueness and overlap of AGN identification 
at different wavelengths and summarize the selection biases of each 
identification method. We further compare the host galaxy properties 
of the AGN in our sample with other studies in the same redshift regime. 
Finally, we discuss our ability to probe the coeval growth of SMBHs and
galaxies at $z\sim2$ with this dataset.

%%%%%%%%%%%%%%%%%%%%%%%%%%%%%%%%%%%%%%%%%%%%%%%%%%%%%%%%%%%%%%%%%%%%%%
%%%%%%%%%%%%%%%%%%%%%%%%%%%%%%%%%%%%%%%%%%%%%%%%%%%%%%%%%%%%%%%%%%%%%%

\subsection{Uniqueness and Overlap of AGN Identified at Different Wavelengths}
\label{sec:depth}

Our sample of 55 AGN at $z\sim$ 1.4--3.8 identified using X-ray, IR and/or optical diagnostics allows us to quantify the uniqueness and overlap of AGN selection at different wavelengths. The numbers of AGN identified at different wavelengths and the overlap between the samples are shown by the Venn diagram in Figure \ref{fig:venn}. As shown in this figure, roughly half of the IR AGN sample and almost half of optical AGN 
sample are {\it not} identified as AGN at the other wavelengths. X-ray AGN identification provides important confirmation of AGN selected at other wavelengths, but in our sample it does not uniquely identify many additional AGN to those identified at MIR and optical wavelengths.

The number of AGN recovered at each wavelength depends on the depth of the observational data available at that wavelength.  
To investigate the differing depths of our observations, we compare the bolometric luminosity for AGN identified at each wavelength. 
We adopt a single bolometric correction at each wavelength.
As mentioned in Section \ref{sec:lx_loiii}, we adopt  $k_\mathrm{X(2-10\; keV)}=25$ and  $k_\mathrm{[OIII]}=600$ respectively for sources with X-ray detections and significant \ox detections. 
At MIR wavelengths we adopt the average bolometric correction from \cite{Richards2006} at 5.8 \um, giving $k_\mathrm{IR}=8$. Although a single bolometric correction is likely an oversimplification, it is sufficient for our purposes to compare
the effective depths of the data at different wavelengths.
The median bolometric luminosities are $L_{bol(X)} = 10^{45.1}$~erg~s$^{-1} $ for sources with X-ray detections, $L_{bol([OIII])} = 10^{45.2}$~erg~s$^{-1}$ for sources with significant \ox detections and $L_{bol(IR)} = 10^{45.3}$~erg~s$^{-1}$ for sources with 5.8 \um \; detections. The similar median bolometric luminosities indicate that our data are reaching similar depths at each wavelength. 
However, there are AGN identified at a single wavelength that are not recovered at other wavelengths. Are these AGN intrinsically less luminous at other wavelengths or is their unique identification due to another observational bias?

While X-ray imaging is a robust method for identifying AGN with hydrogen column densities up to $N_{H}\approx 10^{23-24}$ cm$^{-2}$, X-ray emission cannot penetrate higher column densities and will therefore not identify Compton-thick AGN. In addition, variation of the depth of {\it Chandra} observations in our various fields as well as a changing effective depth within a field results in a non-uniform flux limit  \citep[see, e.g., ][] {mendez2013primus}. 
Therefore, X-ray imaging may miss AGN that are identified at 
other wavelengths.
Furthermore, X-ray selection is not expected to identify many AGN that cannot be recovered at other wavelengths with sufficiently deep data. 
Indeed, in MOSDEF we find that the majority (87\%) of X-ray AGN are also recovered with IR or optical methods.

We find that 75\% of our X-ray AGN are recovered at optical wavelengths.
There are six X-ray AGN that are {\it not} identified at optical wavelengths: 
four of these sources have low S/N optical emission lines that are 
contaminated by sky lines; 
one of these sources is at $z>3$ where \ha and \nii fall beyond the wavelength coverage of MOSFIRE and therefore cannot be placed on the BPT diagram; and one source is on the star-forming sequence in the BPT diagram, indicating that it has a high SFR relative to the AGN luminosity and could therefore not be identified as an optical AGN. We thus conclude that optical AGN selection could identify the majority of X-ray selected AGN and is mainly limited by the quality of the spectroscopic data. However, this method is likely biased against AGN in host galaxies with high SFRs \citep[e.g.][]{coil2015} as sources with higher SFR move towards the star formation locus on the BPT diagram.

Slightly less than half (42\%) of the optical AGN in our sample are not identified at X-ray or IR wavelengths, the majority in fields with relatively shallow X-ray data. 
These differences likely reflect the non-uniform depths probed by the X-ray data, compared to the fairly uniform depth at \ox probed with the MOSDEF spectra. Also, as the \cite{donley2012} selection is very incomplete, these optical AGN are not identified using our IR selection criteria. Thus, optical selection can potentially identify substantial populations of AGN that are missed at other wavelengths.

% IR only
There are 13 IR AGN in our sample that are not selected at X-ray or optical wavelengths. Although these sources have significant \ox fluxes, they cannot be classified as optical AGN for various reasons.  
All of these sources either do not have observations of \ha and \nii 
(typically because they are at $z\gtrsim3$) or have sky line contamination such that their \xbpt ratio cannot be measured.
Based on our upper limits on X-ray luminosities, it appears that these 13 IR AGN are not identified at X-ray wavelengths due to the depths of the available X-ray data; indeed, 11 of these sources are in the COSMOS and AEGIS fields, where we have shallower X-ray data. Thus, IR AGN selection does \emph{not} appear to identify a substantial AGN population (such as heavily obscured sources) that cannot be identified at other wavelengths. 
However, IR selection provides a more uniform depth than X-ray selection and is not affected by the data quality issues that impact optical selection; thus IR selection can be used to improve the completeness of AGN samples. 

\citet{mendez2013primus} reached similar conclusions regarding IR AGN selection using a larger sample of AGN at intermediate redshifts ($z<1.2$), finding that 90\% of IR AGN identified with shallow IR data are detected with sufficiently deep X-ray data. 
As the depth of the IR observations increases, \citet{mendez2013primus} find that the fraction of IR AGN that are {\it not} recovered by X-rays also increases, reflecting the additional IR AGN samples that are identified with extremely deep IR data.
Using deep IR data, \cite{donley2012} find that just 38\% of IR AGN in their sample are recovered at X-rays wavelengths, although as the depth of the X-ray data increases this fraction increases to 52\% \citep[see also][]{Donley2007,hickox2009host}. 
More recently, \cite{Cowley2016} find X-ray counterparts for only $\sim22\%$ of their IR AGN. \cite{Cowley2016} use the \cite{Messias2012} redshift-dependent IR AGN selection criteria; thus the lower fraction of X-ray counterparts in their work could be either due to shallower X-ray data or the different IR selection method used.

\begin{table*}
\caption{ The selection biases of X-ray, IR and optical AGN host galaxies in MOSDEF}
\begin{center}
\renewcommand{\arraystretch}{2}
\begin{tabular}{m{1.3in} m{1.3in} m {1.4in} m {1.1in}}
\hline \hline

Host galaxy property & X-ray AGN & IR AGN & optical AGN  \\
\hline \hline 

Stellar mass  & bias\;towards\; \;\; \; \; \; \;\;\; high mass galaxies & bias\;towards\; \;\; \; \; \;moderate mass galaxies & 
bias\;towards\; \;\; \; \; \;high mass galaxies \\  
%\hline
 
SFR & no bias  & bias\;towards \; \;\; \; \; \; \;\;\;\;\;\;\;\;\;\;\ relatively higher SFR  &bias towards \; \;\; \; \; \; \;relatively lower SFR\\ 
%\hline 

Dust &possible\;bias\;towards \; \;\; \; \; \;  higher dust content &possible bias towards \; \;\; \; \; \;  lower dust content & possible\;bias\;towards \; \;\; \; \; \;  higher dust content  \\ 
\hline 

\end{tabular}
\end{center}
\label{tab:bias}
\end{table*}

%%%%%%%%%%%%%%%%%%%%%%%%%%%%%%%%%%%%%%%%%%%%%%%%%%%%%%%%%%%%%%%%%%%%%%
%%%%%%%%%%%%%%%%%%%%%%%%%%%%%%%%%%%%%%%%%%%%%%%%%%%%%%%%%%%%%%%%%%%%%%

\subsection{AGN Selection Biases}
\label{sec:biases}

As shown above and elsewhere, there are substantial observational selection biases in AGN samples identified at different wavelengths. 
These biases impact the observed properties of the AGN host galaxies identified.

% X-ray AGN
In terms of X-ray identification, 
our results indicate that X-ray selection can identify AGN at low specific accretion rates, which results in a selection bias towards massive host galaxies (see Figures \ref{fig:LOIII_hist} and \ref{fig:sfr_mass}). Indeed, previous studies have extensively shown that AGN identification {\it at any wavelength} is biased against low mass galaxies \citep[e.g.][]{kauffmann2003,xue2010color,aird2012primus}. This bias has also been seen in various studies of X-ray AGN host galaxies \citep[e.g.][]{Alonso2008,aird2012primus,Azadi2015}. In terms of star formation activity of X-ray AGN hosts, although they are mostly located below the main sequence of star formation, their SFR distribution is not significantly different from that of inactive galaxies with a similar mass distribution, as most galaxies at that high stellar mass are also below the main sequence.

% IR AGN
In terms of IR AGN identification, IR AGN selection is biased towards identifying AGN with high specific accretion rates where the AGN IR light dominates over the host galaxy light \citep{mendez2013primus}. This selection bias can result in identifying more luminous AGN in moderately massive host galaxies. Using a larger sample at intermediate redshifts, \cite{mendez2013primus} find that IR AGN selection mainly identifies high X-ray luminosity AGN, while X-ray selection identifies AGN with a wider range of luminosities. Given that most IR AGN in our sample are not detected at X-ray wavelength we cannot make such a comparison here. We further find that within MOSDEF, IR AGN are found in less dusty host galaxies with relatively younger stellar populations and higher SFRs. 
This effect can be understood as related to the stellar mass selection biases for IR AGN: IR AGN are identified in lower stellar mass galaxies (compared to X-ray or optical AGN) that tend to have less dust and younger stellar populations than higher mass galaxies.

% optical AGN
In terms of optical AGN identification, we find that optical selection 
can identify lower accretion rate AGN that may not be recovered at other 
wavelengths. 
Considering the similar \lo distributions for the various AGN in our sample, this trend is driven by the high stellar masses of the optical AGN host galaxies \citep[see also][]{coil2015}, similar to X-ray AGN. 
We further find that optical AGN reside in dusty galaxies with older stellar populations and relatively moderate star formation activity (median log({SFR/SFR${_\mathrm{MS}}$)= -0.87 dex).}
The higher stellar mass of the optical AGN host galaxies leads to a bias towards higher dust content. 
It is also more likely for optical AGN to be identified in galaxies with older stellar populations and lower SFR in the BPT diagram \citep{coil2015}.
We further note that the bias towards more massive host galaxies 
leads to a bias towards higher metallicities, therefore the optical selection method may not be successful in identifying low mass-low metallicity host galaxies \citep[e.g.][]{Groves2006}. We emphasize that although our sample is small, the selection biases of optical AGN against lower mass galaxies with higher SFR has been also reported in studies of optical AGN at lower redshifts with large samples \citep[][]{kauffmann2003,Trump2015}.

Overall, we find that compared to IR and optical AGN selection techniques, 
X-ray identification is the least biased, with only a bias towards high 
stellar mass (but no additional SFR bias).  
We summarize the various selection biases discussed above in Table \ref{tab:bias}.

%%%%%%%%%%%%%%%%%%%%%%%%%%%%%%%%%%%%%%%%%%%%%%%%%%%%%%%%%%%%%%%%%%%%%%

\subsection{MOSDEF Host Galaxy Properties Compared to the Literature }
\label{sec:host_gal}

In this study, we find no significant differences between MOSDEF AGN host galaxies and inactive galaxies of the same stellar mass. 
In particular, we find no significant differences between the star formation activity of AGN host galaxies with the inactive mass-matched galaxies for AGN selected at a given wavelength. 
Thus, after taking into account of the observational selection biases, we find no evidence that AGN activity is preferentially occurring in a particular type of galaxy, although our relatively small sample size may preclude us from identifying weak trends. 

The same result has been seen in some recent studies, e.g. \cite{rosario2015} find a similar SFR distribution in X-ray AGN hosts and mass-matched galaxies at $z\sim2$. \cite{bongiorno2012accreting} also find similar distribution of sSFR for AGN and the inactive galaxies with a slight increase in AGN fraction towards lower sSFR. On the other hand, a number of studies find that AGN are preferentially found in star-forming (main sequence) galaxies at these redshifts. \cite{Azadi2015} perform X-ray sensitivity corrections and find that X-ray AGN are $2-3$ times more likely to be found in galaxies with elevated SFR \citep[see also][]{aird2012primus,santini2012enhanced,harrison2012no,Bernhard2016}. Recently, \cite{Mullaney2015} use \textit{Herschel} and \textit{ALMA} measurements and find that the majority of X-ray AGN are {\it below} the main sequence of star formation, arguing that studies using mean-stacking for SFR measurements can overestimate the level of star formation in the host galaxies.

The MOSDEF sample does not contain a large number of 
quiescent galaxies, which lack high S/N emission lines at the observed wavelengths of the survey \citep{kriek2015}. However, the fraction of our AGN in the quiescent region of UVJ space is similar to the fraction of mass-matched galaxies in that region. With a small number of quiescent galaxies in our sample, the majority of our AGN are in star-forming galaxies which is consistent with the results from other studies at $z\sim2$. However, to robustly determine whether AGN host galaxies lie preferentially below, above, or along the main sequence of star formation will require larger samples for investigations.

\cite{Ellison2016} considered a sample of multi-wavelength identified AGN at $z\sim 0$, and similar to our results find
IR AGN in galaxies with elevated SFR relative to the main sequence and optical AGN in galaxies with lower SFR than the main sequence.  \cite{Cowley2016} performed a similar analysis at $z \lesssim 3$ , and find 
that the specific star formation rate (sSFR, $\frac{SFR}{\mathcal{M}_{*}}$) 
in AGN host galaxies is, on average, higher than in mass-matched galaxies at $z\gtrsim2$ for their IR selected AGN sample. No significant differences in sSFR were found for X-ray or radio AGN at these redshifts.

However, in our sample we find a similar sSFR distribution for IR AGN and mass-matched galaxies. As IR AGN in MOSDEF span a similar range of stellar mass as the IR AGN identified in  \cite{Cowley2016}, the higher sSFR in  \cite{Cowley2016} must be due to higher SFR for IR AGN in their sample. Additionally, our investigation shows that the majority of IR AGN in our sample reside in less dusty star-forming galaxies, while \cite{Cowley2016} at $z\gtrsim2$ find the majority of IR AGN in dusty star-forming galaxies. 
The different SFR and dustiness of IR AGN in our sample compared to \cite{Cowley2016} could be due  to the fact that  \cite{Cowley2016} use a combination of IRAC and 24 \um\; observations for IR AGN identification \citep[see][]{Messias2012}.

We further used the location of AGN host galaxies in UVJ space to investigate their dust properties. The location of galaxies in UVJ space is very sensitive to stellar mass, with lower mass galaxies residing preferentially in the less dusty star-forming region, and more massive star-forming galaxies in the dusty region \citep[e.g.][]{whitaker2011}. While the full AGN sample used here shows a similar distribution as 
mass-matched inactive galaxies in UVJ space, we find that the fraction of 
optical AGN in the dusty star-forming region is higher than their inactive mass-matched galaxies at the $2.9\sigma$ level. However, the X-ray and IR AGN show a very similar behavior to their mass-matched galaxies. Although our optical AGN hosts are predominantly in dusty star-forming 
galaxies, we note that the $A_{V}$ of these optical AGN hosts is not significantly higher than in the mass-matched galaxies or non-optical AGN population. Therefore, the difference in the fraction of optical AGN and their mass-matched galaxies in the UVJ diagram could be a statistical fluctuation, rather than due to any intrinsic difference in dust content of the optical AGN host galaxies.

Overall, the AGN in our sample have very similar physical properties to those of mass-matched inactive galaxies. At these redshifts, a larger sample of both quiescent galaxies and AGN are required to study any potential differences between the SFR or dustiness of AGN hosts and inactive galaxies more robustly.

%%%%%%%%%%%%%%%%%%%%%%%%%%%%%%%%%%%%%%%%%%%%%%%%%%%%%%%%%%%%%%%%%%

\subsection{ Are the Growth of Black holes and the Growth of their Host Galaxies  Correlated at $z\sim2$?}

As discussed in the introduction, the global SMBH accretion rate density 
and SFR density both peak at $z\sim2-3$ 
\citep[e.g.,][]{Aird2015}, which indicates that globally there is a relation between the growth of SMBH and their host galaxies. But the question still remains whether such a correlation exists on the scale of individual host galaxies. Our results indicate that \sfratio and AGN luminosity are not significantly correlated within our sample, using either \lx or \lo as a probe of AGN activity. Why then, given the similarity in the global scaling relations, we do not find a correlation?

Due to their stochastic fueling \citep[e.g.][]{ulrich1997variability,peterson2001variability}, the luminosities of AGN may undergo dramatic changes in a short time scale \citep{keel2012history}, while star formation activity remains stable in the host galaxies over long timescales \citep[e.g.][]{wong2009timescale,hickox2012laboca}. Therefore rapid AGN variability can play an important role in washing out any underlying trend that may exist between \sfratio and AGN luminosity. In fact, studies using average AGN luminosity in bins of SFR, instead of the luminosity of individual AGN, find a positive trend between AGN luminosity and SFR of the host galaxy. \citep[e.g.][]{chen2013correlation,hickox2014black,Azadi2015,dai2015}.

Although the connection between galaxy-wide star formation and AGN activity might be hidden due to the variable nature of AGN, \cite{diamond2012relationship} find evidence of a strong correlation between AGN luminosity and SFR in the circumnuclear regions ($r < 1$ kpc).
Thus, while our results do not show a significant correlation between the large-scale star formation and AGN activity, these phenomena may have an underlying connection through a common gas supply. We note that violent events such as major mergers can provide a gas influx to fuel both AGN activity and star formation. However, the moderate luminosity of AGN in our sample indicate that these sources are generally at lower luminosities than those thought to be triggered by major mergers \citep[e.g.][]{schawinski2012,Treister2012}.

As discussed above in Section \ref{sec:contamination}, AGN and star formation activity in the host galaxy can both contribute to the \ox luminosity. In the local Universe studies have proposed various methods for estimating the contribution from star formation to the \ox emission line \citep[e.g.][]{kauffmann2009,wild2010,tanaka2011}. At the redshifts and depth of the MOSDEF survey only a fraction of 
our AGN can be accurately placed on the BPT diagram, due to 
contamination from sky lines and the lack of spectroscopic wavelength coverage at higher redshifts. Therefore, the commonly-used methods for estimating the star formation contribution to \lo at low redshifts 
cannot be applied to our sample. Although \lo can be boosted by star formation activity, here we do not find a significant correlation between SFR and \lo, which indicates that it is unlikely for the instantaneous star formation rate to be correlated with \lo.

\section{Summary}
\label{sec:summary}

In this paper we use the data from the first two years of the MOSDEF survey to investigate AGN identification and their host galaxies properties at $1.37<z<3.80$, with the majority of our sample at $z\sim2$. We identify 55 AGN using the X-ray imaging data from \textit{Chandra}, mid-IR data from IRAC camera on \textit{Spitzer}, and rest-frame optical spectra from MOSDEF survey. We investigate the selection biases from each identification method and explore the host galaxy properties of these AGN. We further consider the relation between star formation activity and AGN luminosity in our sample. Our main conclusions are as follows:

\begin{itemize}

\item We find that AGN identified at any wavelength are biased against low mass host galaxies; this is an observational selection bias. IR AGN identification has an additional bias against the most massive galaxies. Quantifying the SFR relative to the main sequence and comparing the distributions for IR and optical AGN, we find that IR AGN are primarily identified in galaxies with relatively higher SFRs, while optical AGN are identified in galaxies with relatively lower SFRs ($p$=0.004, at $> 2\sigma$ significance). X-ray selection does not display any bias in the SFR distribution relative to the main sequence. The observational biases in stellar mass can result in biases in terms of the dust content of host galaxies, with IR AGN showing a possible bias towards less dusty host galaxies and optical and X-ray AGN showing a possible bias towards more dusty host galaxies in our sample. 

\item Within the star-forming galaxy population, once stellar mass selection biases are taken into account, we find that AGN reside in galaxies with similar physical properties (SFR, dust content, and stellar age) as inactive galaxies. Therefore we find no evidence of AGN activity in particular types of galaxies, which is consistent with stochastic fueling of AGN in any kind of galaxy, and no strong evidence for AGN feedback.

\item The majority of the AGN in our sample can be identified using optical diagnostics. We find that 75\% of the X-ray AGN in our sample are also identified with optical diagnostics, indicating the reliability of optical AGN selection. However, optical identification is limited by the quality of the spectroscopic data, as optical emission lines in most of the non-optical AGN in our sample at $z\sim2$ are contaminated by night sky lines.

\item  Almost half of the IR AGN in our sample are recovered at X-ray or optical wavelengths. IR imaging provides a more uniform depth than X-ray data and is not affected by the quality of optical spectroscopy; thus IR AGN identification can improve the completeness of AGN samples at $z\sim2$.

\item The relationship between \lx and \lo  in our sample at $z\sim2$ is consistent with the relation of \cite{heckman2005} in the local Universe. Unlike \cite{heckman2005}, who found that the majority of local optical AGN can be recovered at X-ray wavelengths, we find X-ray counterparts for only 50\% of the optical AGN in our sample. This is likely due to the relatively shallower and variable depth of the X-ray data across our fields.

\item We do not find a significant correlation between \sfratio (SFR relative to the main sequence of star formation) and AGN luminosity (using \lx or \lo) in our sample. Although \lo can be boosted by star formation activity in the host galaxy, at $z\sim 2$ we cannot apply correction techniques commonly used at lower redshifts to estimate the SF contamination. 

\end{itemize}

Although the selection biases in our sample are derived from a small number of AGN, they are consistent with results of studies at intermediate redshifts with larger samples. The presence of these selection biases indicates that in order to obtain a more complete AGN census, complementary identification techniques at multiple wavelengths are required. To robustly study AGN host galaxy properties, the selection biases from each identification technique should be taken into account.

\acknowledgements
We thank the MOSFIRE instrument team for building this powerful instrument, and for taking data for us during their commissioning runs. This work would not have been possible without the 3D-HST collaboration, who provided us the spectroscopic and photometric catalogs used to select our targets and to derive stellar population parameters. Based on observations made with the NASA/ESA Hubble Space Telescope, which is operated by the Association of Universities for Research in Astronomy, Inc., under NASA contract NAS 5-26555. These observations are associated with programs 12177, 12328, 12060-12064, 12440- 12445, 13056. Funding for the MOSDEF survey is provided by NSF AAG grants AST-1312780, 1312547, 1312764, and 1313171 and grant AR-13907 from the Space Telescope Science Institute. A.L.C. acknowledges support from NSF CAREER award AST-1055081. N.A.R. is supported by an Alfred P. Sloan Research Fellowship. JA acknowledges support from ERC Advanced Grant FEEDBACK 340442. The data presented herein were obtained at the W. M. Keck Observatory, which is operated as a scientific partnership among the California Institute of Technology, the University of California and the National Aeronautics and Space Administration. The Observatory was made possible by the generous financial support of the W. M. Keck Foundation. The authors wish to recognize and acknowledge the very significant cultural role and reverence that the summit of Mauna Kea has always had within the indigenous Hawaiian community. We are most fortunate to have the opportunity to conduct observations from this mountain.

\bibliographystyle{apjurl}
\bibliography{references.bib}

\maketitle

\end{document}